\documentclass[a4paper,11pt]{article}
\pdfoutput=1

\usepackage{jcappub}

\title{A Universal Bound on Excitations of Heavy Fields during Inflation}

\author[a]{Thorsten Battefeld}
\author[a,b]{ and R. C. Freitas}
\affiliation[a]{Institute for Astrophysics, University of Goettingen, Friedrich Hund Platz 1, D-37077 Goettingen, Germany}
\affiliation[b]{Universidade Federal do Esp\' \i rito Santo, Centro de Ci\^encias Exatas, Departamento de F\' \i sica, Av. Fernando Ferrari, 514 Campus de Goiabeiras, CEP 29075-910, Vit\'oria, Esp\' \i rito Santo, Brazil}

\emailAdd{tbattefe(at)gmail.com, \\ rodolfo.camargo(at)pq.cnpq.br}

\abstract{We discuss a universal bound on {\emph{any}} excitation of heavy fields during inflation: the ratio of the heavy field's energy density to the one driving inflation must be less than the maximally allowed relative amplitude of oscillations in the power-spectrum ($\rho_{\mathrm h}/\rho_{\mathrm I} \lesssim 0.01 $ according to PLANCK). This bound can be traced back to the sudden change of the equation of state parameter across the excitation event.

We employ a sudden transition approximation at the perturbed level, which has been used before in different settings; we check its validity by comparison to the full multi-field result in a concrete case study involving a sudden mass change of an inflaton. }


\begin{document}
\maketitle
\flushbottom

\section{Introduction}
\label{sec:intro}

How strongly can a heavy field, $m_{\mathrm h}\gg H$, be excited (i.e., displaced from its vacuum expectation value) during inflation without spoiling inflationary predictions? We show that only a small fraction of the inflationary energy density can be transferred to a heavy field if oscillations in the power-spectrum in excess of current bounds are to be avoided ($\rho_{\mathrm h}/\rho_{\mathrm I} \lesssim 0.01 $ if relative oscillations should contribute no more than one percent). This bound is universal, i.e., independent of the excitation mechanism and the inflationary model. Its physical origin is the sudden change of the equation of state parameter across the transition that is present for any excitation. After a pedagogical derivation of this bound, which has been derived previously in different frameworks, we discus its applicability and implications; we follow with a novel case study to highlight its universality. 
  
Current observations of the cosmic microwave background radiation \cite{Ade:2013uln,Ade:2013ydc,Ade:2013zuv,Ade:2014xna} are consistent with the predictions of simple, single field models of inflation (adiabatic, Gaussian perturbations with a slightly red power-spectrum and observable gravitational waves, see \cite{Bassett:2005xm,Baumann:2009ds,Martin:2013tda,Baumann:2014nda} for reviews), yet, a priori, most models in string theory are of the multi-field type \cite{Baumann:2014nda}: even if a hierarchy of masses can justify the use of an effective single-field model \cite{Assassi:2013gxa}, heavy fields are present and can get excited during inflation. 
For example, if the inflaton is identified with the light direction along a channel of the potential, any turn\footnote{The same effect can be achieved in a straight channel with a non-flat field space metric.} of this trough would lead to deviations of the trajectory from the minimum \cite{Burgess:2012dz,Cespedes:2013rda}, and thus a displacement of the heavy field from its vacuum expectation value -- the heavy field would be excited. If the bending is mild, the effects at the perturbed level can be captured by an effective single-field theory\footnote{Concerns raised in \cite{Avgoustidis:2012yc} regarding the applicability of an EFT were primarily due to incompatible definitions of the term EFT, see \cite{Burgess:2012dz} for a clarifying discussion.} (EFT) with a time dependent speed of sound \cite{Tolley:2009fg,Achucarro:2010jv,Achucarro:2010da,Achucarro:2012sm,Achucarro:2012yr,Burgess:2012dz,Achucarro:2012fd,Achucarro:2014msa} and clear tell-tale sings in observables (e.g.~a ringing pattern in the power-spectrum at the scale corresponding to the Hubble radius when the turn was encountered \cite{Shiu:2011qw,Cespedes:2012hu,Gao:2012uq,Konieczka:2014zja}); see \cite{Cremonini:2010ua,Peterson:2011yt,Behbahani:2011it,Shiu:2011qw,Chen:2012ge,Pi:2012gf,Gwyn:2012mw,Gao:2013zga,Emami:2013lma,Castillo:2013sfa} for related work. For sharper bends, the EFT fails and different multi-field analytic or numerical techniques \cite{Gao:2012uq,Gao:2013ota,Noumi:2013cfa,Konieczka:2014zja}  have to be used at the perturbed level. 

How much energy can be transferred to the heavy field while remaining consistent with current observations? The PLANCK satellite considerably improved limits on oscillatory features in the power-spectrum \cite{Ade:2013uln}, leading to constraints on variations of the speed of sound within the EFT \cite{Ade:2013uln,Achucarro:2014msa} or model parameters in concrete setups, such as monodromy inflation \cite{Easther:2013kla}. As a rule of thumb, corrections at the percent level are not ruled out, but their addition does not improve the fit at a statistically significant level \cite{Easther:2013kla}\footnote{For particular oscillatory shapes, amplitudes of up to ten percent can be consistent with the data \cite{Chen:2014joa}; we use one percent as a conservative upper limit in this paper, since essentially any correction below that level (regardless of placement/frequency/shape) is not ruled out. 
}.

Before proceeding, we should be more precise regarding the excitation of a heavy field: we consider homogeneous and isotropic excitations as opposed to the (potentially inhomogeneous) quantum mechanical production of field quanta \cite{Bagger:1997dv,Watson:2004aq}. Within the EFT  \cite{Achucarro:2010jv,Achucarro:2010da,Achucarro:2012sm,Achucarro:2012yr,Burgess:2012dz,Achucarro:2012fd,Achucarro:2014msa}, it is the former, i.e., the deviation of the background trajectory from the minimum of the valley, that affects curvature perturbations. The latter is relevant in models of trapped inflation \cite{Kofman:2004yc,Green:2009ds,Silverstein:2008sg,Battefeld:2010sw,Battefeld:2013bfl}
or modulated trapping  \cite{Langlois:2009jp,Battefeld:2011yj,D'Amico:2013iaa} among other applications: a heavy field becomes temporarily light  so that particle production similar to the one during preheating  \cite{Kofman:1997yn,Battefeld:2006cn,Battefeld:2012wa} can take place. Such heavy field excitations affect the background and perturbations (computing the latter is subtle, since back-reaction is included at the level of the equations of motion and not the Lagrangian, see \cite{Green:2009ds}). If adiabaticity is strongly violated during the excitation event, e.g.~due to a sharp turn, particle production is present in addition to the classical effects discussed here, see \cite{Konieczka:2014zja}.

If a sharp turn is present, rapid oscillations in the heavy direction result, whose repercussions onto perturbations can be challenging to compute. In \cite{Chen:2008wn,Chen:2010xka,Chen:2011zf,Chen:2011tu,Chen:2012ja,Battefeld:2013xka,Chen:2014joa} the oscillations at the background level are kept, but perturbations are truncated to the light field. This approach enables a straightforward computation of correlation functions in the in-in formalism, leading to additional oscillatory features caused by \emph{resonances} which could act as a fingerprint of the background cosmology (i.e., inflationary v.s.~bouncing cosmologies) if they were observed. However, the truncation is generically not viable to compute this effect, since it ignores the coupling to the heavy field's perturbations; for example, the necessary condition needed to truncate perturbations derived in \cite{Battefeld:2013xka} is not satisfied for some case studies in \cite{Chen:2011zf,Chen:2012ja}; other bounds are stated in  
\cite{Burgess:2012dz,Shiu:2011qw}. As explained in detail in \cite{Gao:2012uq,Gao:2013ota,Noumi:2013cfa}, the amplitude of the signal is suppressed by a factor of $H^2/m_{\mathrm{h}}^2$ in comparison to the one in \cite{Chen:2010xka,Chen:2011zf,Chen:2011tu,Chen:2012ja,Battefeld:2013xka}, since neglected terms lead to cancellations\footnote{The computation of the bi-spectrum is more subtle: since the cancellation is not complete, the resulting amplitude in the full analysis remains at the same order of magnitude \cite{Noumi:2013cfa}.} in integrals for the power-spectrum (the shape/frequency and location of the signal is unaffected). Considering derivative interactions between the light and heavy field can boost the signal again \cite{Saito:2012pd,Saito:2013aqa,Kobayashi:2012kc,Chen:2014joa}. These oscillations due to resonances are distinct from the above-mentioned sinusoidal ringing pattern.

Often, the process of exciting the heavy field is not retained in analytic treatments, see e.g.~\cite{Chen:2010xka,Chen:2011zf,Chen:2011tu,Chen:2012ja,Battefeld:2013xka}; further, keeping the coupling of perturbations is challenging and only possible in certain cases \cite{Gao:2012uq,Gao:2013ota} (see also \cite{Noumi:2012vr,Mizuno:2014jja}); thus, a numerical solution to the full multi-field set-up as in \cite{Gao:2012uq,Noumi:2013cfa} is usually necessary if high accuracy is desired: in \cite{Gao:2012uq,Gao:2013ota} turns were investigated in detail by different methods, highlighting the limitations of analytic techniques (EFT or truncation). Unfortunately, in many cases the physical process responsible for oscillations in the power-spectrum is obscured in the final results by the necessarily advanced computational techniques.

Our goal is to reinvestigate a universal, analytic order of magnitude estimate for the expected leading order corrections to the power-spectrum that retains the crucial sinusoidal oscillations (ringing pattern) at scales corresponding to modes that crossed the Hubble radius at the time of the excitation event. To this end, we focus on a sudden change of one field's mass as a concrete example of an excitation event, which provides a complementary mechanism to excite a heavy field in comparison to the commonly discussed turns. We do not consider resonance effects. 

Such an estimate is possible since any excitation causes a small, rapid change of the equation of state parameter $w=p/\rho$.  This change leads to mode mixing at the perturbed level (the Bogoliubov coefficient $\beta$ becomes non-zero, corresponding to particle production) directly leading to oscillations in the power-spectrum. This  process is retained if perturbations are truncated to the light sector and the subsequent oscillations of the heavy field are cut out. It is present whenever the equation of state changes quickly: Joy et.~al.~\cite{Joy:2007na,Joy:2008qd} computed analytically such oscillations for a jump in the mass during single field inflation,
following earlier work in \cite{Starobinsky:1992ts,Adams:2001vc} on steps in the potential. In \cite{Battefeld:2010rf,Battefeld:2010vr}, it was computed in the multi-field framework of staggered inflation \cite{Battefeld:2008py,Battefeld:2008qg}, where fields start to decay during inflation; it is also present if a sharp turn is encountered during inflation or during multi-field open inflation \cite{GarciaBellido:1997te,Sugimura:2011tk,Battefeld:2013xka} \footnote{Generically, a tunnelling event does not position fields onto the inflationary attractor in a valley \cite{Sugimura:2011tk}, providing yet another, natural mechanism to excite heavy fields.}. See \cite{Firouzjahi:2014fda,Firouzjahi:2010ga,Namjoo:2012xs} for related work.

Evidently, this effect has been rediscovered several times and it is indeed {\emph{universal}}, as mentioned in the original work by Joy et al. \cite{Joy:2007na} (see also \cite{Chen:2011zf}): by construction, the energy in a heavy field red-shifts fast, leading to a sudden change of the inflationary equation of state parameter\footnote{One needs to compare $w$ before the transition to its value after the heavy field's energy density became unimportant; the transient (larger amplitude) oscillations in $w$ caused by oscillations in the heavy field are irrelevant for this effect (but crucial for resonance effects).}; therefore, whenever oscillations in a heavy field are excited during inflation, be it by a turn, a non-flat field space metric, a kink in the potential, a tunnelling event or any other mechanism, a ringing in the power-spectrum results.

We compute the amplitude of these oscillations in the power-spectrum in a simple two field inflationary model: initially, both fields have the same mass, but one of them encounters a kink in the potential, turning it instantaneously into a heavy field, see Sec.~\ref{sec:sec1}. We follow \cite{Joy:2007na,Battefeld:2010rf,Battefeld:2010vr} to compute the resulting power-spectrum in the sudden transition approximation, see Sec.~\ref{sec:subsec1.1}. This result has been derived in the framework of staggered inflation \cite{Battefeld:2010rf} following work by Joy et al.~\cite{Joy:2007na}, (see \cite{Chen:2011zf} for a derivation based on the in-in formalism). We provide this straightforward derivation for pedagogical reasons before we discuss the general implications in Sec.~\ref{discussion1}.

To show the applicability of the resulting estimate, we compute perturbations in the full two-field setup in Sec.~\ref{sec:twofielsetup} for comparison, in line with the analysis in \cite{Gao:2012uq,Gao:2013ota}. We find that the sudden transition approximation provides indeed a surprisingly good estimate for the amplitude and frequency of oscillations in the power-spectrum, see Sec.~\ref{discussion2}.  Thus, the main process leading to such a ringing pattern is just the sudden change of the equation of state parameter. 

We would like to reiterate that both aspects, the techniques used in this paper as well as the result based on the sudden transition approximation, are not new. However, the broad applicability of this simple, physically intuitive result appears to be little known outside of a small expert community, which motivated us to provide this case study; our chosen mechanism to excite a heavy field is complementary to the commonly used ones involving turns in the potential valley. We hope our work will aid cosmologists that are less familiar with advanced computational techniques for multi-field cosmological perturbations in the presence of non-trivial inflationary background solutions: the sudden transition approximation enables a quick back of the envelope estimate to decide whether or not a full multi-field analysis of perturbations is warranted in a given inflationary model. If the latter is called for, for example to compare predictions in a concrete model with data sets, the techniques used in e.g.~\cite{Gao:2012uq,Gao:2013ota,Noumi:2013cfa,Achucarro:2014msa} should be used, some of which we employ in this paper to put the estimate to the test.

\section{Background}
\label{sec:sec1}

Consider two canonically normalized scalar fields with the action
\begin{equation}
   \label{eq:action}
	 S = \int{d^4\sqrt{-g}\left( \frac{M_{\mathrm{Pl}}^2}{2}R -\frac{1}{2}g^{\mu\nu}\sum_{i=1}^{2}{\partial_\mu\phi_{i}\partial_\nu\phi_{i}} + V\left(\phi_{1},\phi_{2}\right) \right)}  
\,, 
\end{equation}
and the potential
\begin{equation}
   V\left(\phi_{1},\phi_{2}\right) = \sum_{i=1}^{2}{V_{i}}(\phi_i) \,.\label{def:potential}
\end{equation}
Since we are interested in a simple inflationary model for which one of the fields becomes heavy, we take
\begin{eqnarray}
   V_{i}^-\left( \phi_{i-}\right) & = & \frac{1}{2}m_{i-}^2\phi_{i-}^2 \quad , \quad t\leq t_{*} \quad , \\
	 V_{i}^+\left( \phi_{i+}\right) & = & 
	    \begin{cases}
			   \frac{1}{2}m_{1+}^2(\phi_{1+}-\tilde{\phi}_{1+})^2 \\
				 \frac{1}{2}m_{2+}^2\phi_{2+}^2
			\end{cases}
			\quad , \quad t \geq t_{*} \,,\label{defphitildeh}
\end{eqnarray}
where $t_*$ denotes the transition time at which $\phi_1$ acquires a heavy mass  
\begin{equation}
   m_{\mathrm l}\equiv m_{1-} \equiv m_{2-} \equiv m_{2+} \ll m_{1+}\equiv m_{\mathrm h} \, .
\end{equation}
We impose $t_*$ to lie within the observational window, i.e., around sixty e-folds before the end of inflation. Here and in the following the subscripts $_\pm$ denote the time just before/after the transition. We model the transition as instantaneous, which is a good approximation as long as the time required for the change in mass is much smaller than the Hubble time. $\tilde{\phi}_{1+}$ is chosen such that the potential energy is continuous, i.e., $V_1^+=V_1^-$. In the following we replace $\left\{1,2\right\}\Leftrightarrow \left\{h,l\right\}$, while keeping in mind that $\phi_1$ has a heavy mass  for $t>t_*$ only.

As a background we take a flat, homogeneous and isotropic universe with the Friedmann-Lemaitre-Robertson Walker (FLRW) line element
\begin{equation}
   ds^2=dt^2-a(t)^2d\vec{x}^2 \quad ,
\end{equation}
where $a(t)$ is the scale factor and $t$ is cosmic time. The Friedmann equations for the Hubble factor $H=\dot{a}/a$ and the Klein-Gordon equations are
\begin{eqnarray}
    && 3 H^2 = M_{\mathrm{Pl}}^{-2}\sum_{i}{\left(\frac{\dot{\phi}^{2}_{i}}{2}+V_{i}\right)} \quad , \quad
	   \dot{H} = - M_{\mathrm{Pl}}^{-2}\sum_{i}{\frac{\dot{\phi}^{2}_{i}}{2}} \quad , \\
		 &&\ddot{\phi}_{i}+3H_{}\dot{\phi}_{i}+V_{i}^\prime = 0 \, ,
\end{eqnarray}
where we introduced the shorthand notation $V_i^\prime\equiv \partial_{\phi_{i}}V_i$ and the reduced Planck mass 
\begin{eqnarray}
M_{\mathrm{Pl}}^{-2}\equiv 8\pi G\,.
\end{eqnarray}
As we will see in Sec.~\ref{discussion1}, oscillations in the power-spectrum will be of order
\begin{eqnarray}
\mathcal{\varepsilon} \equiv \frac{\rho_{\mathrm h}}{\rho_{\mathrm l}}\bigg|_{t_*}\,, 
\end{eqnarray}
where 
\begin{eqnarray}
\rho_i=V_i+\frac{1}{2}\dot{\phi}_i^2\,.
\end{eqnarray}
Since the amplitude of oscillations in the power-spectrum is bounded to be less than one percent by current observations \cite{Ade:2013uln,Easther:2013kla}, we impose $\varepsilon \lesssim \mathcal{O}(10^{-2})\ll 1$ in the following. This small parameter should not be confused with the slow-roll parameters
\begin{eqnarray}
   \epsilon & \equiv & -\frac{\dot{H}}{H^2} = \sum_{i}{\frac{M_{\mathrm{Pl}}^{-2}}{2}\left(\frac{\dot{\phi}_{i}}{H}\right)^2}\equiv \sum_i\epsilon_i \quad , \quad \epsilon_{ij}=\sqrt{\epsilon_{i}\epsilon_{j}} 
   \,, \\
	 \delta_{i} & \equiv & - \frac{\ddot{\phi}_{i}}{H\dot{\phi}_{i}} , \quad 
	 \delta_{i} + \delta_{j} = \epsilon - \frac{\dot{\epsilon}_{ij}}{2H\epsilon_{ij}} , \quad \eta_{ij} \equiv M_{\mathrm{Pl}}^2 \frac{\partial^{2}_{\phi_i\phi_j}V}{V} \,.
\end{eqnarray}
For the inflationary phase to be of the slow-roll type, we require $\epsilon\ll 1$ throughout, so that we can approximate
\begin{equation}
   \label{eq:friedmann1}
   3H^2 \simeq M_{\mathrm{Pl}}^{-2}\sum_i V_{i} \,.
\end{equation}

\subsection{Before the Transition, $t \leq t_{*}$}

\begin{figure}[tb]
\begin{center}
\includegraphics[width=0.65\columnwidth]{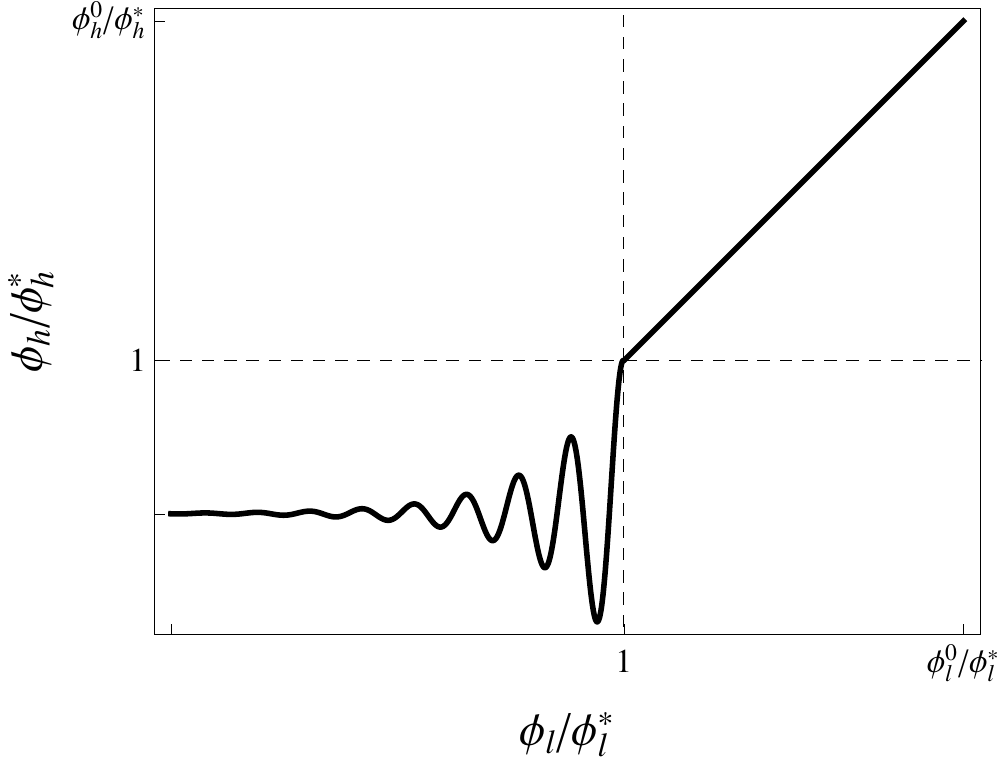} 
\end{center}
\caption{Schematic field evolution covering the transition at $t_*$ of $\phi_{\mathrm h}$ from a light field to a heavy one, as given by the solutions in (\ref{eq:sol_hfield-}) and (\ref{eq:sol_hfield+}).  
The corresponding potential in (\ref{def:potential}) is continuous at the transition $\vec{\phi}^*\equiv (\phi_{\mathrm l}^*,\phi_{\mathrm l}^*$), but the second derivative makes a jump from $\partial^2 V/\partial \phi_{\mathrm h}^2|_-=m_{\mathrm l}^2$ to $\partial^2 V/\partial \phi_{\mathrm h}^2|_+=m_{\mathrm h}^2$ at ${\vec{\phi}}^*$.
}
\label{fig.phase_space}
\end{figure}

Since both fields are initially light, we demand $\delta_{i}\ll 1$ before the transition, leading to 
\begin{equation}
   \label{eq:KG1}
   3H\dot{\phi}_{i}+V_{i}^\prime \simeq 0 
\end{equation}
and 
\begin{equation}
	\epsilon \simeq \sum_{i}{\frac{M_{\mathrm{Pl}}^{2}}{2}\left(\frac{V_{i}^\prime}{3H^2}\right)^2} 
\end{equation}
to leading order in the slow-roll approximation. 
The energy ratio at the transition satisfies
\begin{eqnarray}
\varepsilon \simeq  \frac{\phi_{\mathrm h}^{*2}}{\phi_{\mathrm l}^{*2}}+\frac{1}{3}\epsilon_{\mathrm h} \ll 1 \,; \label{energycondition}
\end{eqnarray}
the slow-roll conditions combined with this energy condition 
 lead to constraints on the field values
\begin{equation}
   \phi_{\mathrm h}^*\ll \phi_{\mathrm l}^* \quad , \quad \phi_{\mathrm l}^* > M_{\mathrm{Pl}} \,,
\end{equation}
so that $m_{\mathrm l}\ll H$. 
Since we want to place $t_*$ in the observational window, we choose 
\begin{eqnarray}
\phi_{\mathrm l}^* \equiv 15 M_{\mathrm{Pl}}
\end{eqnarray}
throughout this paper, which leads to around sixty e-folds of slow-roll inflation driven by $\phi_{\mathrm l}$ for $t>t_*$. 
Solving (\ref{eq:KG1}) in the slow-roll limit leads to
\begin{eqnarray}
   \label{eq:sol_lfield-}
   \phi_{\mathrm l} & \approx & \phi_{\mathrm l}^{*}\left[1+\sqrt{\frac{2}{3}}\left(\frac{M_{\mathrm{Pl}}}{\phi_{\mathrm l}^{*}}\right)m_{\mathrm l}\left(t_{*}-t\right)\right] \, , \\
	\label{eq:sol_hfield-}
	 \phi_{\mathrm h} & \approx & \phi_{\mathrm h}^{*}\left[1+\sqrt{\frac{2}{3}}\left(\frac{M_{\mathrm{Pl}}}{\phi_{\mathrm l}^{*}}\right)m_{\mathrm l}\left(t_{*}-t\right)\right] \, ,
\end{eqnarray}
which are the well known slow-roll solutions with constant speed.
Since 
\begin{eqnarray}
\epsilon_{\mathrm h} &\simeq& 2 M_{\mathrm{Pl}}^2 \frac{\phi_{\mathrm h}^{*2}}{\phi_{\mathrm l}^{*4}}\,,\\
\epsilon_{\mathrm l} &\simeq& 2 M_{\mathrm{Pl}}^2 \frac{1}{\phi_{\mathrm l}^{*2}}
\end{eqnarray}
and $\phi_{\mathrm h}^*\ll \phi_{\mathrm l}^*$,  it follows that
\begin{eqnarray}
\epsilon_{\mathrm h} \ll \epsilon_{\mathrm l}\,,
\end{eqnarray}
i.e., we have to treat $\epsilon_{\mathrm h}$ as a second order quantity. Consequently, the energy condition in (\ref{energycondition}) simplifies to
\begin{eqnarray}
\varepsilon\simeq \frac{\phi_{\mathrm h}^{*2}}{\phi_{\mathrm l}^{*2}} \ll 1\,.\label{varepsilonbeforetransition}
\end{eqnarray}

\subsection{After the Transition, $t \geq t_{*}$}

Once $\phi_{\mathrm h}$ becomes heavy, we focus on cases with $\epsilon\ll 1$ and $\delta_{\mathrm l}\ll 1$, i.e., cases for which the light field is rolling slowly and dominates the energy density, so that the Hubble factor evolves equally slowly. However,  $\phi_{\mathrm h}$ speeds up and breaks slow-roll ($\delta_{\mathrm h}\gg 1$) for $m_{\mathrm h}\gg H$. Introducing 
\begin{eqnarray}
\epsilon_{\mathrm m}\equiv \frac{m_{\mathrm l}^2}{m_{\mathrm h}^2}\,,
\end{eqnarray}
the condition on the heavy field's mass becomes
\begin{eqnarray}
\epsilon_{\mathrm m}\ll 3\epsilon_{\mathrm l}\,,
\end{eqnarray}
so that $(H/m_{\mathrm h})^2$ can be treated as a second order parameter if compared to e.g.~$\epsilon_{\mathrm l}$ and $\varepsilon$.
Under these conditions the background equations are
\begin{equation}
   3H^2 \simeq M_{\mathrm{Pl}}^{-2}V_{\mathrm l} \quad , \quad  3H\dot{\phi}_{\mathrm l}+V_{\mathrm l}^\prime \simeq 0 \quad , \quad \ddot{\phi}_{\mathrm h}+3H\dot{\phi}_{\mathrm h}+V_{\mathrm h}^\prime = 0 \,,
\end{equation}
which are solved by
\begin{eqnarray}
  \label{eq:sol_lfield+}
   \phi_{\mathrm l} & \simeq & \phi_{\mathrm l}^{*}\left[1-\sqrt{\frac{2}{3}}\left(\frac{M_{\mathrm{Pl}}}{\phi_{\mathrm l}^{*}}\right)m_{\mathrm l}\left(t-t_{*}\right)\right] \quad , \\
	\label{eq:sol_hfield+}
	 \phi_{\mathrm h}-\tilde{\phi}_{\mathrm h} & \simeq & C \left(\frac{a}{a_{*}}\right)^{-3/2}\sin{\left[m_{\mathrm h}(t-t_{*}) + \gamma\right]} \quad ,
\end{eqnarray}
where $a(t_{*})=a_{*}$, $\tilde{\phi}_{\mathrm h}$, $C$ and $\gamma$ are constants of integration. 

\begin{figure}[tb]
\begin{center}
\includegraphics[width=0.8\columnwidth]{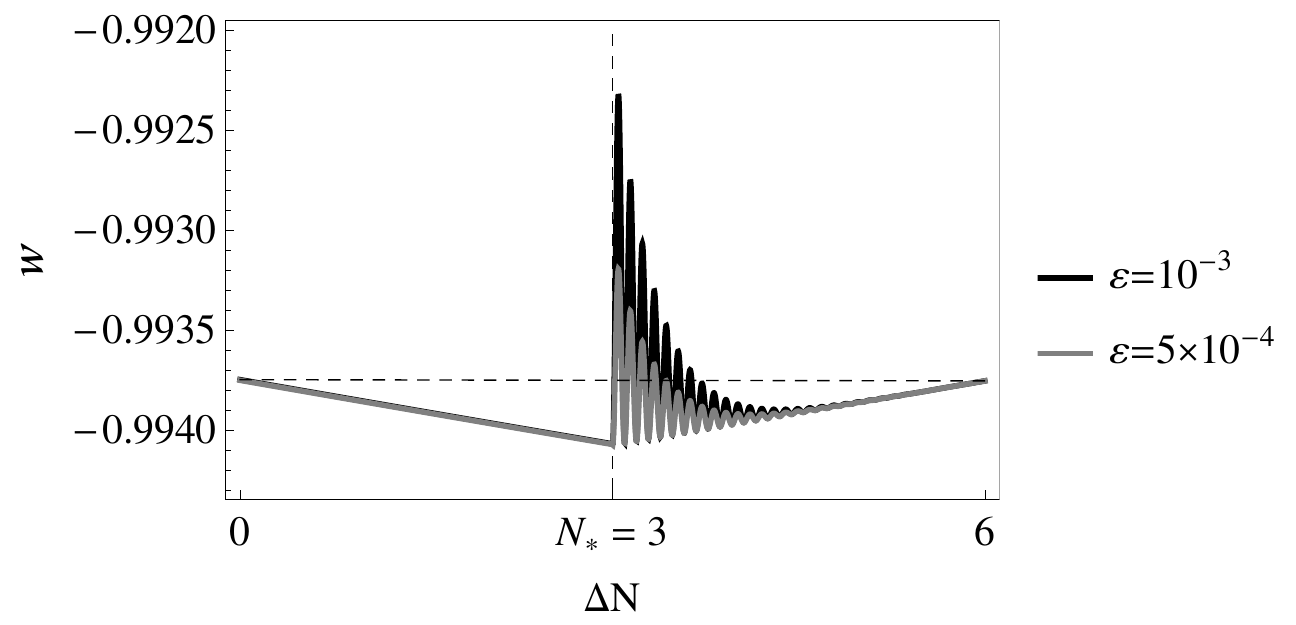} 
\end{center}
\caption{The equation of state parameter in (\ref{defeosp}) as a function of the number of e-folds ($\Delta N$) for $m_{\mathrm l}/m_{\mathrm h}=\sqrt{\epsilon_{\mathrm m}} = 5\times 10^{-3}$ and two different energy ratios; the oscillations die down within a few e-folds. The entire plot covers six efolds and $N_*\equiv 3$ indicates the transition point.}
\label{fig.EoS_param}
\end{figure}

At the transition we need to impose the continuity of the scale factor and  the Hubble parameter
\begin{equation}
  [a]^{\pm} = 0\quad ,\quad  [H]^{\pm} = 0 \,,
\end{equation}
as dictated by the Israel/Deruelle-Mukhanov junction conditions \cite{Israel:1966rt,Deruelle:1995kd,Martin:1997zd}, which require the continuity of the extrinsic curvature and induced metric at the hyper-surface separating the two inflationary regimes.
 For the transition under consideration, this hyper-surface is set by the field value $\phi_{\mathrm{h}}$, similar to hybrid inflation  \cite{Linde:1993cn, Copeland:1994vg}. Here, we introduced the notation
\begin{eqnarray}
 [F]^{\pm}\equiv F(t^+_{*})-F(t^-_{*})\,,
 \end{eqnarray} 
 to designate the change of a quantity over the transition. Since  fields are not interacting or decaying, we further impose the continuity of $\phi_i$,  their respective potentials as well as their kinetic energies, i.e., $[\rho_i]^\pm=0$ and $[\dot{\phi}_i]^{\pm} = 0$.
Plugging the solutions (\ref{eq:sol_lfield-}) and (\ref{eq:sol_hfield-}) as well as (\ref{eq:sol_lfield+}) and (\ref{eq:sol_hfield+}) into these matching conditions, we arrive after some algebra at 
\begin{eqnarray}
\label{defgammafull}\cot{\gamma} &=& \frac{3}{2}\frac{H}{m_{\mathrm h}}+\sqrt{\frac{2}{3}\left(\frac{M_{\mathrm{Pl}}}{\phi_{\mathrm l}^*}\right)^2}\simeq \frac{1}{\sqrt{3\epsilon_{\mathrm l}}}\left(\frac{3}{2}\sqrt{\epsilon_{\mathrm m}}+\epsilon_{\mathrm l}\right)\ll 1\, , \label{gamma}
\end{eqnarray}
so that we may approximate
\begin{eqnarray}
\gamma \approx \frac{\pi}{2}- \frac{1}{\sqrt{3\epsilon_{\mathrm l}}}\left(\epsilon_{\mathrm l}+\frac{3}{2}\sqrt{\epsilon_{\mathrm m}}\right)\equiv \frac{\pi}{2}+\delta\gamma\,, \label{defgamma}
\end{eqnarray}
after Taylor expanding $\cot \gamma$ around $\pi/2$. Similarly, we get
\begin{eqnarray}
   C &=& \left(\frac{m_{\mathrm l}}{m_{\mathrm h}}\right)\frac{\phi_{\mathrm h}^*}{|\sin{\gamma}|}
  \approx \sqrt{\epsilon_{\mathrm m}}(1+\delta\gamma)\phi_{\mathrm h}^*\approx \sqrt{\epsilon_{\mathrm m}}\phi_{\mathrm h}^*\,,
   \end{eqnarray}
   and
   \begin{eqnarray}
   \tilde{\phi}_{\mathrm h} &=& \phi_{\mathrm h}^*\left(1-\frac{m_{\mathrm l}}{m_{\mathrm h}}\right)= \phi_{\mathrm h}^*\left(1-\sqrt{\epsilon_{\mathrm m}}\right)\,.
\end{eqnarray}
where $\tilde{\phi}_{\mathrm h}$ was introduced in (\ref{defphitildeh}). We use the approximate expression for $\gamma$ in analytic estimates, but keep the full one in plots.
The ratio of energies in (\ref{energycondition}) remains
\begin{eqnarray}
\mathcal{\varepsilon} \equiv \frac{\rho_{\mathrm h}^*}{\rho_{\mathrm l}^*} \simeq \frac{\phi_{\mathrm h}^{*2}}{\phi_{\mathrm l}^{*2}}\,,
\end{eqnarray}
since $[\rho_i]^\pm =0$. 

While we treated the transition from $m_{\mathrm l} \rightarrow m_{\mathrm h}$ for $\phi_{\mathrm h}$ as instantaneous, the resulting damped oscillations of the heavy field in (\ref{eq:sol_hfield+}) take a few Hubble times to die off. Since we assumed quadratic potentials, the energy density of the heavy field redshifts as $a^{-3}$; thus, the amplitude of the heavy field decreases rapidly. In Fig.~\ref{fig.phase_space} we plot schematically the evolution of the fields.
 Evidently, our setup provides a viable mechanism to excite oscillations in a heavy field without interrupting inflation or disturbing the background solution of the light field.

The initial oscillation amplitude of the heavy field is set by the energy ratio of the fields,
\begin{eqnarray}
\Xi \equiv \phi_{\mathrm h}^*-\tilde{\phi}_{\mathrm h}=\frac{m_{\mathrm l}}{m_{\mathrm h}}\phi_{\mathrm l}^*\sqrt{\varepsilon}=\phi_{\mathrm h}^*\sqrt{\epsilon_{\mathrm m}}\approx C\,.\label{defXi}
\end{eqnarray}
Due to the oscillations in $\phi_{\mathrm h}$, the equation of state (EoS) parameter
\begin{equation}
   1+w=\frac{\dot{\phi}_{\mathrm l}^2+\dot{\phi}_{\mathrm h}^2}{3M_{\mathrm{Pl}}^2H^2} \,,\label{defeosp}
\end{equation}
inherits an oscillatory component, see Fig.~\ref{fig.EoS_param} for two examples. These oscillations are equally short lived.

\subsection{Sudden Transition Approximation}
\label{sec:subsec1.1}
 If we ignore the transient oscillation in $w$, we arrive at the \emph{sudden transition approximation}, which retains the main physical effect responsible for the ringing pattern in the power-spectrum, while enabling analytic estimates that are independent of the particular excitation mechanism of a heavy field, see Sec.~\ref{discussion1}. We check the validity of this approximation by using the full, two field setup in Sec.~\ref{sec:twofielsetup}. 

In the sudden transition approximation we ignore the contribution of the heavy field after the transition entirely, so that 
\begin{equation}
   1+w_{+}\approx \frac{\dot{\phi}_{\mathrm l}^2}{3M_{\mathrm{Pl}}^2H^2}\simeq \frac{2}{3}\epsilon_{\mathrm l} \,,
\end{equation}
while
\begin{eqnarray}
  1+w_{-}\simeq  \frac{2}{3}(\epsilon_{\mathrm l}+\epsilon_{\mathrm h}) \,.
\end{eqnarray}
Hence, the fractional change in $w$ becomes
\begin{equation}
   \frac{\Delta w}{1+w_-} \equiv \frac{w_- - w_+}{1+w_-} \approx \frac{\varepsilon}{1+\varepsilon}\approx \varepsilon \, ,\label{jumpineosparameter}
\end{equation}
where we used $\epsilon_{\mathrm h}/\epsilon_{\mathrm l}\simeq \varepsilon$. Since the two slow-roll regimes are smoothly glued together in the sudden transition approximation, the Hubble parameter, scale factor and conformal time are continuous.

\section{Perturbations}
\label{sec:sec2}
We are interested in computing the power-spectrum of the curvature perturbation $\mathcal{R}$ after the transition when the transient oscillations of the heavy field died down. To this end, we provide an estimate based on the sudden transition approximation of Sec.~\ref{sec:subsec1.1}, as well as a more refined but less transparent analysis based on retaining perturbations of both fields.  The computations in this section follow primarily \cite{Taruya:1997iv,Gordon:2000hv,Joy:2007na,Battefeld:2010rf,Battefeld:2010vr}. For a brief review of cosmological perturbation theory see e.g.~\cite{Bassett:2005xm}.

The general perturbed line element in a FLRW Universe including only scalar degrees of freedom is
\begin{equation}
   ds^2=(1-2A)dt^2+2aB_{,i}dx^idt-a^2[(1-2\psi)\delta_{ij}+2E_{,ij}]dx^idx^j \,.
\end{equation}
Two of the four degrees of freedom are pure gauge. Gauge invariant metric perturbations can be defined via
\begin{eqnarray}
    \Phi & = & A+ (aB-a^2\dot{E}) \, , \label{Bardeenpotential}\\
		\Psi & = & \psi - H(aB-a^2\dot{E}) \,,
\end{eqnarray}
which coincide in the absence of anisotropic stress, $\Phi=\Psi$, which is the case for us. Perturbing the scalar fields
\begin{eqnarray}
\phi_i({\bf{x}},t)\equiv \phi_i(t)+\delta \phi_i({\bf{x}},t)\,,
\end{eqnarray}
it is straightforward to derive the coupled equations of motion for these gauge dependent field perturbations in Fourier space \cite{Gordon:2000hv,Taruya:1997iv}
\begin{equation}
   \ddot{\delta \phi}_{i}+3H\dot{\delta \phi}_{i}+\frac{k^2}{a^2}\delta \phi_{i}+\sum_{j}{(\partial^{2}_{\phi_i\phi_j}V)\delta \phi_{j}}=-2(\partial_{\phi_i}V)A+\dot{\phi}_{i}\left[\dot{A}+3\dot{\psi}+\frac{k^2}{a^2}(a^2\dot{E}-aB)\right] \quad .
\end{equation}
These equations simplify after introducing the gauge invariant Sasaki-Mukhanov variables
\begin{equation}
    Q_i=\delta \phi_i+\frac{\dot{\phi}_i}{H}\psi \, ,
\end{equation}
which coincide with the field perturbations in the spatially flat gauge ($\psi = 0$), to 
\cite{Gordon:2000hv}
\begin{equation}
   \ddot{Q}_i+3H\dot{Q}_i+\frac{k^2}{a^2}Q_i+\sum_j{\left(\partial^{2}_{\phi_i\phi_j}V-\frac{M_{\mathrm{Pl}}^{-2}}{a^3}\left(\frac{a^3}{H}\dot{\phi}_i\dot{\phi}_j\right)^{.}\right)Q_j}=0 \, .\label{eomQi}
\end{equation}

\subsection{Before the Transition, $t\leq t_{*}$}

The equations of motion in (\ref{eomQi}) simplify further by introducing the rescaled Sasaki-Mukhanov variable $u_i \equiv aQ_i$ and employing conformal time $dt = a d\tau$.
While both fields are rolling slowly and the universe is inflating, conformal time is given by $\tau\simeq-(1+\epsilon)/(aH)$ to first order in the slow-roll parameters. As a consequence, (\ref{eomQi}) simplifies to
\begin{equation}
   u_{i}^{''}+\left(k^2-\frac{2}{\tau^2}\right)u_{i}=-\frac{1}{\tau^2}\sum_j{M_{ij}u_{j}} \quad ,
\end{equation}
where a prime denotes a derivative with respect to the conformal time and the interaction matrix reads
\begin{equation}
   M_{ij} \simeq 3\left(\eta_{ij}-2\epsilon_{ij}-\epsilon\delta_{ij}\right)\label{defintmatrix1} \,, 
\end{equation} 
to first order in slow-roll.

These two equations can be decoupled by performing a rotation such that $M$ is diagonal \cite{Byrnes:2006fr}. This rotation matrix is
\begin{equation}
   \label{eq:rotation}
   U = \left(
	 \begin{array}{c c}
		  \cos \theta & -\sin \theta \\
			\sin \theta & \cos \theta
	 \end{array}
	 \right) \quad ,
\end{equation}
with the rotation angle $\theta$ given by
\begin{eqnarray}
   \tan 2\theta &=& \frac{2M_{\mathrm{hl}}}{M_{\mathrm{hh}}-M_{\mathrm{ll}}} \,.
\end{eqnarray}
Pluging the slow-roll parameters into (\ref{defintmatrix1}), we can express the interaction matrix in terms of $\phi_{\mathrm l}^*$ and the energy ratio $\varepsilon$ in (\ref{varepsilonbeforetransition}),
\begin{equation}
   M^- \simeq 
	 -6\epsilon_{\mathrm l} (1-2\varepsilon)\left(
	 \begin{array}{c c}
		  \varepsilon & \sqrt{\varepsilon} \\
			\sqrt{\varepsilon} & 1 
	 \end{array}
	 \right) \quad ,
\end{equation}
so that
\begin{eqnarray}
   \tan 2\theta &\simeq & -2\frac{\sqrt{\varepsilon}}{(1-\varepsilon)} \simeq -2 \sqrt{\varepsilon}\,,
\end{eqnarray}

Applying the rotation matrix (\ref{eq:rotation}) to the interaction matrix leads to the desired diagonalization,
\begin{equation}
   U^{T}MU= \left(
   \begin{array}{c c}
	    \lambda_{1}^- &  0 \\
			0 & \lambda_{2}^-
	 \end{array} 
	 \right) \quad ,
\end{equation}
with the eigenvalues
\begin{equation}
   \lambda_{h,l}^- = \frac{1}{2}\left[M_{\mathrm{hh}}+M_{\mathrm{ll}}\pm|M_{\mathrm{hh}}-M_{\mathrm{ll}}|\sqrt{1+(\tan 2\theta)^2}\right] \, .
\end{equation}
To leading order in slow-roll and for $\varepsilon \ll 1$ we find 
\begin{eqnarray}
   \lambda_{\mathrm h}^- &\simeq& \mathcal{O}(\epsilon_{\mathrm l}\epsilon) \,,\label{lambdah+}\\
   \lambda_{\mathrm l}^- &\simeq& 
   \simeq -6\epsilon_{\mathrm l}+\mathcal{O}(\epsilon_{\mathrm l}\epsilon)
   \,.\label{lambdal-}
\end{eqnarray}
The decoupled perturbation variables $v_{i}$ are given by
\begin{equation}
   u_{i} = \sum_{j}{U_{ij}v_{j}} \quad,
	 \label{uv}
\end{equation}
and satisfy
\begin{equation}
   v_{i}''+\left(k^2-\frac{(2-\lambda_i^-)}{\tau^2}\right)v_{i} \approx 0 \,.
\end{equation}
Imposing the Bunch Davies vacuum state $v_{i}\rightarrow e^{-\dot{\imath}k\tau}/\sqrt{2k}$ for  $\tau\rightarrow -\infty$ we find the well known solutions
\begin{equation}
   v_{i}^- = \frac{\sqrt{\pi}}{2}e^{\dot{\imath}(\mu_{i}^-+1/2)\pi/2}\sqrt{-\tau}\mathcal{H}_{\mu_{i}^-}^{(1)}(-k\tau)e_{i}(\vec{k}) \, ,
	 \label{solv-}
\end{equation}
where $\mathcal{H}_{\mu_{i}^-}^{(1)}(-k\tau)$ is the Hankel function of the first kind of order
\begin{equation}
   \mu_{i}^- \equiv \sqrt{\frac{9}{4}-\lambda_{i}^-}  
\end{equation}
 and $e_{i}(\vec{k})$ are independent unit Gaussian random variables satisfying
\begin{equation}
   \left\langle e_{i}(\vec{k})\right\rangle = 0 \quad , \quad \left\langle e_{i}(\vec{k})e_{j}(\vec{k}')^{*}\right\rangle = \delta_{ij}\delta^3(\vec{k}-\vec{k}')\,.
\end{equation}
We were quite verbose in this section, since the same steps will be used below in the less simple setup after the transition. Note that the deviation of $\mu_{\mathrm l}$ from $3/2$ is set by the slow-roll parameters, which determines the deviation of the power-spectrum from scale invariance, i.e., $n_s-1$. While predictions of $n_s$ and the tensor to scalar ratio $r$ based on a quadratic potential are marginally ruled out by Planck at the $2\sigma$-level \cite{Ade:2013uln,Ade:2013zuv}, they fit well with the preliminary detection of gravitational waves by BICEP2 \cite{Ade:2014xna}. We use a quadratic potential throughout this article, but it is straightforward to generalize our results to more complicated forms of $V_{\mathrm l}$ should a different scalar spectral index or tensor to scalar ratio be desired.

\subsection{After the Transition $t\geq t_{*}$}
For $t\geq t_{*}$ conformal time can still be expressed via $\tau\approx -(1+\epsilon)/(aH)$ to first order in slow-roll, since the inflationary regime is not interrupted. The corresponding equations of motion for $u_i$ remain formally  unaltered as well,
\begin{equation}
   u_{i}^{''}+\left(k^2-\frac{2}{\tau^2}\right)u_{i}=-\frac{1}{\tau^2}\sum_j{M_{ij}u_{j}} \, ,
\end{equation}
but the interaction matrix 
\begin{equation}
   M_{ij} \equiv 3(\eta_{ij}+2\epsilon_{ij}-\epsilon\delta_{ij})+6\epsilon_{ij}\left(\frac{V'_i}{3H\dot{\phi}_i}+\frac{V_j^\prime}{3H\dot{\phi}_j}\right) 
\end{equation} 
differs from (\ref{defintmatrix1}), since the heavy field oscillates, see equation (\ref{eq:sol_hfield+}). Consequently, the slow-roll approximation can not be applied for $V_{\mathrm h}^\prime/(3H\dot{\phi}_{\mathrm h})$. Nevertheless,  the rotation angle needed to instantaneously diagonolize the interaction matrix can be kept small for all relevant scales after the transition for suitable choices of model parameters.
To leading order in small parameters we find
\begin{eqnarray}
   M^+_{\mathrm{ll}} & \simeq & -6\epsilon_{\mathrm l} -9\varepsilon F_{\mathrm{osc}}^2 \quad, \\
	 M^+_{\mathrm{hl}} & \simeq & 6\epsilon_{\mathrm l}\sqrt{\frac{\varepsilon}{\epsilon_{\mathrm m}}}\tilde{F}_{\mathrm{osc}} \quad, \\
	 M^+_{\mathrm{hh}} & \simeq & \frac{m_{\mathrm h}^2}{H^2}-3\epsilon_{\mathrm l}+9\varepsilon F_{\mathrm{osc}}^2+12\varepsilon \frac{m_{\mathrm h}}{H}F_{\mathrm{osc}}\tilde{F}_{\mathrm{osc}} \quad ,
\end{eqnarray}
where
\begin{eqnarray}
   F_{\mathrm{osc}} & \equiv & \left(\frac{a}{a_*}\right)^{-3/2}\left(\frac{\cos{\left(m_{\mathrm h}(t-t_*)+\gamma\right)}-\frac{3}{2}\frac{H}{m_{\mathrm h}}\sin{\left(m_{\mathrm h}(t-t_*)+\gamma\right)} }{\sin{\gamma}}\right) \, , \\
   &\approx&-\left(\frac{a_*}{a}\right)^{3/2}\sin(m_{\mathrm h}(t-t_*))\,,\\
	 \tilde{F}_{\mathrm{osc}} & \equiv & \left(\frac{a}{a_*}\right)^{-3/2}\left(\frac{ \sin{\left(m_{\mathrm h}(t-t_*)+\gamma\right)} }{\sin{\gamma}}\right)\approx \left(\frac{a_*}{a}\right)^{3/2}\cos(m_{\mathrm h}(t-t_*)) \,,
\end{eqnarray}
and we approximated $\gamma\approx \pi/2$ according to (\ref{defgamma}) while keeping only the zeroth order in $H/m_{\mathrm h}$.
After some algebra, we find to leading order in small parameters
\begin{eqnarray}
   \tan{2\theta_+} &\simeq& \frac{4\sqrt{\epsilon_{\mathrm m}\varepsilon}\tilde{F}_{\mathrm{osc}}}{1+\epsilon_{\mathrm m}+12\varepsilon \frac{H}{m_{\mathrm h}}F_{\mathrm{osc}}\tilde{F}_{\mathrm{osc}}} \,,\\
   &\approx& 4\sqrt{\epsilon_{\mathrm m} \varepsilon}\tilde{F}_{\mathrm{osc}}\ll 1\,,
\end{eqnarray}
In the small angle approximation, we get the eigenvalues
\begin{eqnarray}
   \lambda_{\mathrm h}^{+} & \approx & \frac{m_{\mathrm h}^2}{H^2} \approx \frac{3\epsilon_{\mathrm l}}{\epsilon_{\mathrm m}}\gg 1\, , \\
	 \lambda_{\mathrm l}^{+} & \approx &  -6\epsilon_{\mathrm l}-9\varepsilon F_{\mathrm{osc}}^2\,. 
\end{eqnarray}
Because $\phi_{\mathrm h}$ became heavy, $\lambda_{\mathrm h}$ picks up a large contribution compared to (\ref{lambdah+}), while $\lambda_{\mathrm l}$ acquires a transient oscillatory contribution compared to (\ref{lambdal-}).
The solution for the modes $v_i$ are again given by Hankel functions
\begin{equation}
   v_{i}^+ = \frac{\sqrt{\pi}}{2}\sqrt{-\tau_{+}}\left[A_{i}(k)\mathcal{H}_{\mu_{i}^+}^{(1)}(-k\tau_{+})+B_{i}(k)\mathcal{H}_{\mu_{i}^+}^{(2)}(-k\tau_{+})\right] \,,
	 \label{solv+}
\end{equation}
where 
\begin{eqnarray}
\mu_i^+\equiv \sqrt{\frac{9}{4}-\lambda_i^+}\,.
\end{eqnarray}
Since $u_i=aQ_i$ and $u_i=\sum_{j}{U_{ij}v_{j}}$, each Sasaki-Mukhanov variable has contributions from the perturbations of the light and heavy field. The amplitude $A_i$ and $B_i$ need to be determined via matching to the pre-transition perturbations, see Sec.~\ref{sec:subsec2.1}.

\subsection{Sudden Transition Approximation}
In the sudden transition approximation introduced in Sec.~\ref{sec:subsec1.1} we ignore the contribution of the heavy field at the perturbed level entirely; the only impact of $\phi_{\mathrm h}$ is to induce a change in the equation of state parameter, since $\rho_{\mathrm{h}}$ decreases rapidly for $t>t_*$. Such a truncation can be justified if $\varepsilon \ll 1$, so that the excitation amplitude after the transition is not too big. As shown in \cite{Battefeld:2013xka}, a necessary (but not sufficient) conditions is
\begin{eqnarray}
\frac{\Xi}{M_{\mathrm{Pl}}} \ll \frac{m_{\mathrm l}}{m_{\mathrm h}}=\sqrt{\epsilon_{\mathrm m}}\label{conditiononXi}\,.
\end{eqnarray}
Since $\Xi=\phi_{\mathrm l}^*\sqrt{\varepsilon\epsilon_{\mathrm m}}= 15 M_{\mathrm{Pl}}\sqrt{\varepsilon\epsilon_{\mathrm m}}$ according to (\ref{defXi}), the above condition is satisfied for $\varepsilon \ll 0.01$. Since corrections to the power-spectrum  computed in Sec.~\ref{Sec:powerspectrumsudden} are of order $\varepsilon$, the sudden transition approximation can be used to recover corrections to $\mathcal{P}_\mathcal{R}$ up to $1\%$ at most; incidentally, the latter is the upper bound set by current observations \cite{Ade:2013uln,Easther:2013kla}. 

In this approximation, the equation of motion for perturbations in the light field (\ref{eomQi}) is
\begin{equation}
   \ddot{Q_{\mathrm l}}+3H\dot{Q_{\mathrm l}}+\frac{k^2}{a^2}Q_{\mathrm l}+\left(V_{\mathrm l}^{\prime \prime}-\frac{M_{\mathrm{Pl}}^{-2}}{a^3}\left(\frac{a^3}{H}\dot{\phi}_{\mathrm l}^2\right)^{.}\right)Q_{\mathrm l}=0 \,,
\end{equation}
with the solutions
\begin{eqnarray}
   Q_{\mathrm l}^- & = & \frac{\sqrt{\pi}}{2}e^{\dot{\imath}(\nu_{\mathrm l}^-+1/2)\pi/2}\sqrt{-\tau}\mathcal{H}_{\nu_{\mathrm l}^-}^{(1)}(-k\tau)e_{\mathrm l}(\vec{k}) \,, \label{Ql-}\\
	 Q_{\mathrm l}^+ & = & \frac{\sqrt{\pi}}{2}\sqrt{-\tau}\left[\alpha(k)\mathcal{H}_{\nu_{\mathrm l}^+}^{(1)}(-k\tau)+\beta(k)\mathcal{H}_{\nu_{\mathrm l}^+}^{(2)}(-k\tau)\right]e_{\mathrm l}(\vec{k})\label{Ql+} \,,
\end{eqnarray}
where $\nu_{\mathrm l}^+ = \nu_{\mathrm l}^- \approx \sqrt{9/4+6\epsilon_{\mathrm l}}$. The two Bogoliubov coefficients $\alpha$ and $\beta$ need to be determined by appropriately matching the two solutions, as we shall discuss in the next section.

\section{Matching Conditions at $t_*$ and the Power-spectrum}
\label{sec:subsec2.1}

Treating the transition from $m_{\mathrm l} \rightarrow m_{\mathrm h}$ as instantaneous, we use again the Israel/Deruelle-Mukhanov matching conditions \cite{Israel:1966rt,Deruelle:1995kd,Martin:1997zd} to relate the perturbations before and after the transition. Since the transition takes place at a uniform $\phi_{\mathrm h}$ surface, the conditions reduce to the continuity of  \cite{Battefeld:2010rf}
\begin{equation}
    \label{eq:matching1}
		[\mathcal{R}]^\pm=0 \quad , \quad [\Phi]^\pm=0 \, ,
\end{equation}
where $\Phi$ is the Bardeen potential in (\ref{Bardeenpotential}), and the comoving curvature perturbation is defined as (see \cite{Bassett:2005xm} for a review)
\begin{equation}
   \mathcal{R} = \Phi +\frac{2}{3(1+\omega)}\left(\frac{\Phi'}{aH}+\Phi\right) \, ,
\end{equation}
which can be related to the Sasaki-Mukhanov variables via
\begin{equation}
   \label{eq:curvature}
   \mathcal{R}=\frac{H}{\left(\sum_j{\dot{\phi}_j^2}\right)}\left(\sum_i{\dot{\phi}_iQ_i}\right) \, .
\end{equation}
We can also write $\Phi$ in terms of the Sasaki-Mukhanov variables \cite{Battefeld:2010rf}
\begin{equation}
    \label{eq:scalarpotential}
   -k^2\Phi = \frac{a^2}{2M_{\mathrm{Pl}}^2}\sum_i{\left(\dot{Q}_i\dot{\phi}_i+Q_i\left(\frac{\partial V}{\partial\phi_i}+\frac{3}{2}(1-\omega)\dot{\phi}_iH\right)\right)} \,,
\end{equation}
so that we can translate the matching conditions for $\mathcal{R}$ and $\Phi$ into conditions for the $Q_i$.

\subsection{Sudden Transition Approximation}
Ignoring the contribution of the heavy field, (\ref{eq:curvature}) and (\ref{eq:scalarpotential}) simplify to
\begin{eqnarray}
   -k^2\Phi & \approx & a^2\frac{\dot{\phi}_{\mathrm l}\dot{Q}_{\mathrm l}}{2M_{\mathrm{Pl}}^2} \,, \\
	 \mathcal{R} & \approx & \frac{\dot{\phi}_{\mathrm l}Q_{\mathrm l}}{3(1+\omega)M_{\mathrm{Pl}}^2H} \, ,
\end{eqnarray}
so that the matching conditions in (\ref{eq:matching1}) become
\begin{equation}
   \left[\dot{Q}_{\mathrm l}\right]^{\pm}=0 \quad , \quad \left[\frac{Q_{\mathrm l}}{1+\omega}\right]^{\pm} \, .
	 \label{Deruelle-Mukhanov}
\end{equation}
Thus, even though the heavy field's perturbations are ignored, the jump in the equation of state parameter in (\ref{jumpineosparameter}), caused by the ``vanishing'' of the heavy field after its oscillations died down, leads to a non-trivial matching of solutions.
Plugging the solutions in (\ref{Ql-}) and (\ref{Ql+}) into the matching conditions, we can calculate the Bogoliubov coefficients to
\begin{eqnarray}
   \alpha & = & -\frac{\pi}{8}\dot{\imath}e^{\dot{\imath}(\nu_{\mathrm l}+1/2)\pi/2}\left\{3\frac{\Delta w}{1+w_-}\mathcal{H}_{\nu_{\mathrm l}}^{(2)}\mathcal{H}_{\nu_{\mathrm l}}^{(1)}+2x\left[\mathcal{H}_{\nu_{\mathrm l}}^{(2)}\frac{d\mathcal{H}_{\nu_{\mathrm l}}^{(1)}}{dx}-\frac{1+w_+}{1+w_-}\mathcal{H}_{\nu_{\mathrm l}}^{(1)}\frac{d\mathcal{H}_{\nu_{\mathrm l}}^{(2)}}{dx}\right]\right\} \, , \\
	  \beta & = & \frac{\pi}{8}\dot{\imath}e^{\dot{\imath}(\nu_{\mathrm l}+1/2)\pi/2}\left\{3\frac{\Delta w}{1+w_-}\mathcal{H}_{\nu_{\mathrm l}}^{(1)}\mathcal{H}_{\nu_{\mathrm l}}^{(1)}+2x\left[\mathcal{H}_{\nu_{\mathrm l}}^{(1)}\frac{d\mathcal{H}_{\nu_{\mathrm l}}^{(1)}}{dx}-\frac{1+w_+}{1+w_-}\mathcal{H}_{\nu_{\mathrm l}}^{(1)}\frac{d\mathcal{H}_{\nu_{\mathrm l}}^{(1)}}{dx}\right]\right\} \, , 
\end{eqnarray}
where $\nu_{\mathrm l}^+\approx \nu_{\mathrm l}^-=\sqrt{9/4+6\epsilon_{\mathrm l}}$ and all Hankel functions depend on the dimensionless variable 
\begin{eqnarray} 
x \equiv -k\tau_* \,.
\end{eqnarray}
 Since the details of the excitation mechanism, and in fact the effect of the heavy field at the perturbed level, are cut out in the sudden transition approximation, we deduce that the above Bogoliubov coefficients are a generic feature. Not surprisingly, the same coefficents were found in \cite{Battefeld:2010rf} where inflatons in multi-field brane/anti-brane inflation decayed during inflation as the associated pairs of branes annihilate, leaving some radiation as decay products (see also \cite{Battefeld:2010vr,Firouzjahi:2010ga,Namjoo:2012xs,Firouzjahi:2014fda}). 

\subsubsection{The Power-spectrum \label{Sec:powerspectrumsudden}}
As a consequence of $\beta\neq 0$, the power-spectrum $\mathcal{P}_{\mathcal{R}}$ of the curvature perturbation $\mathcal{R}$, which is defined by
\begin{equation}
   \mathcal{P}_{\mathcal{R}}\delta^{(3)}(\vec{k}-\vec{k'}) = \frac{4 \pi k^3}{(2 \pi)^3}\left\langle \mathcal{R}(\vec{k'})^*\mathcal{R}(\vec{k})\right\rangle \quad ,
\end{equation}
inherits an oscillatory correction on sub-Horizon scales, $-k\tau \gg 1$,
\begin{eqnarray}
\Delta\mathcal{P}_{\mathcal{R}}+1\equiv  \frac{\mathcal{P}_{\mathcal{R}}}{\mathcal{P}^{\mathrm{SR}}_{\mathcal{R}}} & = &\cos^2{\left[x+\left(\nu_{\mathrm l}+\frac{1}{2}\right)\frac{\pi}{4}\right]}+\left(\frac{1+w_+}{1+w_-}\right)^2\sin^2{\left[x+\left(\nu_{\mathrm l}+\frac{1}{2}\right)\frac{\pi}{4}\right]} \label{powerspectrumcorrectionapprox_0} \\
	 	& \approx &  1 - 2\varepsilon \sin^2{\left(-x+\frac{\pi}{2}\right)} \,, \label{powerspectrumcorrectionapprox}
\end{eqnarray}
where we expanded the Hankel functions in the large argument limit and used trigonometric identities, see \cite{Battefeld:2010rf} for details as well as the full result in terms of Bessel functions and an approximation for super-horizon scales. Here, $\varepsilon=\rho_{\mathrm h}/\rho_{\mathrm l}$ should not be confused with the slow-roll parameter $\epsilon$.

 $\mathcal{P}^{\mathrm{SR}}_{\mathcal{R}}$ is the slow-roll power-spectrum of a single light field in a quadratic potential with the usual scalar spectral index,
\begin{equation}
   \mathcal{P}_{\mathcal{R}}^{\mathrm{SR}} \equiv \left(\frac{H^2}{2\pi \dot{\phi}_{\mathrm l}}\frac{\Gamma(\mu_{\mathrm l}^+)}{\Gamma(3/2)}\right)^2\left|-\frac{k\tau_+}{2}\right|^{3-2\mu_{\mathrm l}^+} \,,
\end{equation} 
see e.g.~\cite{Bassett:2005xm}.

\subsubsection{Discussion \label{discussion1}}
The oscillations in (\ref{powerspectrumcorrectionapprox}) continue indefinitely, since the excitation event has not been resolved; the finite duration of the latter leads to a suppression for large $x$; as a consequence, modes with oscillation frequency much higher than the inverse excitation duration are unaffected, see Sec.~\ref{discussion2}. The correction in (\ref{powerspectrumcorrectionapprox_0}) has been derived previously  in different contexts, see e.g.~\cite{Battefeld:2010rf,Chen:2011zf}, and it is valid for any sudden change of the equation of state parameter during slow-roll inflation as long as several conditions are met:
\begin{enumerate}
\item Fluctuations in the light field start out in the Bunch Davies vacuum.
\item Fields have canonical kinetic terms.  If this condition is relaxed, sub-leading resonance effects can become important, see e.g.~the discussion of turns in two-field DBI inflation in \cite{Mizuno:2014jja} or the tachyonic falling of a derivatively coupled field in \cite{Chen:2014joa}.
\item The coupling of light fields to the degrees of freedom responsible for $\Delta w$ can be ignored. This includes the condition on the heavy field's amplitude in (\ref{conditiononXi}), which is satisfied for $\varepsilon\lesssim 0.01$ (consistent with current observational constraints).
\item Quantum particle production can be ignored. The latter can become important for excitation events that strongly violate adiabaticity, see \cite{Konieczka:2014zja}.  
\item Inflationary slow-roll is not interrupted by the transition, i.e., the slow-roll conditions for the light field are satisfied and $-\dot{H}/H^2$ remains small.
\end{enumerate}
Under these conditions we expect the ringing pattern in  (\ref{powerspectrumcorrectionapprox_0}) to provide an approximation for {\it any} excitation mechanism of heavy fields during inflation.

Given the apparent broad applicability of (\ref{powerspectrumcorrectionapprox}), let us discuss its implications: the amplitude of oscillations is set by the energy-ratio $\varepsilon=\rho_{\mathrm h}/\rho_{\mathrm l}$ at the transition. The amplitude of these oscillations in often bigger than the amplitude caused by resonances of the (here unresolved) oscillations of background quantities with perturbations, as computed in \cite{Chen:2010xka} and follow-up papers (the amplitude in 
\cite{Chen:2010xka,Chen:2011zf,Chen:2011tu,Chen:2012ja,Battefeld:2013xka} was corrected in \cite{Gao:2012uq,Gao:2013ota}, reducing its value by a factor of $(H/m_{\mathrm h})^2$). Thus, the mere presence of an excitation and its associated change in $w$ often provides the leading order contribution; it is thus crucial to begin the analysis before the heavy field got excited.  A similar statement holds for the EFT developed in \cite{Achucarro:2010jv,Achucarro:2010da,Achucarro:2012sm,Achucarro:2012yr,Burgess:2012dz,Achucarro:2012fd,Achucarro:2014msa} or numerical methods.

If the excitation event is a sharp turn, $\rho_{\mathrm{h}}$ increases due to the transfer of kinetic energy. During slow-roll inflation the latter is slow-roll suppressed so that $\varepsilon \lesssim \epsilon$. 
As a consequence, we expect $\Delta\mathcal{P}_{\mathcal{R}}\sim \varepsilon \lesssim \epsilon \sim \mathcal{O}(10^{-2})$, that is,  corrections at the percent level at most. This estimate is consistent with the numerical studies of a sharp turn in two-field inflation \cite{Konieczka:2014zja}. It should be noted that particle production during the non-adiabatic process of exciting a heavy field can lead to order one corrections in $\Delta\mathcal{P}_{\mathcal{R}}$ \cite{Konieczka:2014zja}. We do not discuss these effects here.

Oscillations at the percent level  can marginally improve a fit of the power-spectrum to CMBR observations by Planck \cite{Ade:2013uln,Easther:2013kla,Chen:2014joa}, but the cost of adding three new parameters (location, frequency and amplitude) render the improvements statistically insignificant. Thus, taking a conservative point of view, we deduce that oscillations are absent in the current data leading to an experimental upper bound\footnote{It might be interesting to estimate the bi-spectrum caused by a sudden change in the equation of state parameter in a similar manner; however, any comparison with data should include cross-correlating signals in the bi-spectrum with the ones in the power-spectrum to improve the statistical significance and pinpoint the physical origin of oscillations. See e.g.~\cite{Mizuno:2014jja} (oscillations due to resonances in DBI inflation), \cite{Achucarro:2012fd,Achucarro:2013cva} (mild turn, EFT) or \cite{Gong:2014spa,Achucarro:2014msa}. Our estimate is not sufficient for this task, since the phase of oscillations is not recovered properly.} of
\begin{eqnarray}
\boxed{\varepsilon = \frac{\rho_{\mathrm h}}{\rho_{\mathrm l}}\lesssim 0.01\,.} \label{conditiononenergy}
\end{eqnarray}
 For quadratic potentials the above translates to an upper bound on the excitation amplitude of the heavy field of
 \begin{eqnarray}
 \Xi \simeq \frac{m_{\mathrm l}}{m_{\mathrm h}} \phi_{\mathrm l}^* \sqrt{\varepsilon} \leq \sqrt{\epsilon_{\mathrm m}} 1.5 M_{\mathrm{Pl}} \,. \label{conditiononamplitude}
 \end{eqnarray}
Thus, any excitations of heavy fields in the observational window are strongly constrained, the more so the heavier the field under consideration. At this point we observe that the condition in (\ref{conditiononXi}) on $\Xi$, which is needed to truncate perturbations to the light sector, is valid up until the current observational upper limit is approached.

We put the sudden transition approximation to the test in the next section, where we provide a two field analysis at the perturbed level during the transition caused by the sudden change of one field's mass. This example provides a complementary case study to the commonly discussed turns during inflation that we briefly mentioned above. This computation is the main new aspect of this article and, while somewhat cumbersome, provides additional support for the applicability of  (\ref{powerspectrumcorrectionapprox_0}).

\subsection{The Full two Field Setup \label{sec:twofielsetup}}
\begin{figure}[tb]
   \begin{center}
      \includegraphics[width=0.70\columnwidth]{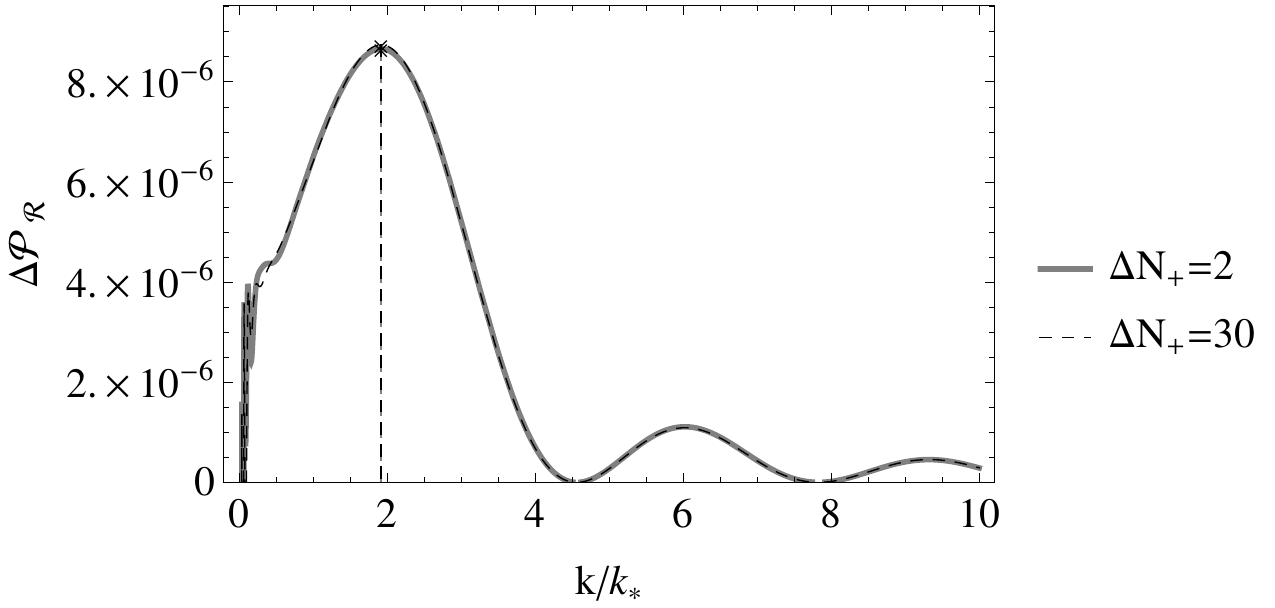}
	    \caption{The correction to the power-spectrum in the full two field setup with $\varepsilon = 10^{-6}$, $\epsilon_{\mathrm m} = 10^{-3}$ and $\phi_{\mathrm l}^*=15 M_{\mathrm{Pl}}$, plotted shortly after the transition and substantially later. The first peak is located at $k/k_* \approx 1.92$ with an amplitude of $8.66 \times 10^{-6}$ for the the gray curve ($\Delta N_+=2$) and $8.74 \times 10^{-6}$ for the the black dashed curve ($\Delta N_+=30$). After a few e-folds an adiabatic regime is entered and (\ref{fullpowerspectrum}) can be used.}
	 \label{delta_p_1}
	 \end{center}		
\end{figure}
We wish to compute the Bogoliubov coefficients while resolving the transition and keeping both fields at the perturbed level. Since the potential and kinetic energy of each field are continuous, the equation of state parameter is continuous too,
\begin{eqnarray}
[w]^\pm=0\,.
\end{eqnarray}
Thus, only the discontinuity in $m_{\mathrm l}\rightarrow m_{\mathrm h}$ for $\phi_{\mathrm h}$ can bring about a non-trivial matching condition. 

Although the suddenly heavy field induces a sharp turn (see Figure \ref{fig.phase_space}), background quantities remain smooth
\begin{eqnarray}
[a]^\pm=[H]^\pm=[\rho_i]^\pm=[V_i]^\pm=0\,.
\end{eqnarray}
To impose the Isreal/Deruelle-Mukhanov matching conditions, we impose (\ref{eq:matching1}) for each summand in (\ref{eq:curvature}) and (\ref{eq:scalarpotential}) individually, as in \cite{Battefeld:2010vr} (fields do not interact during the instantaneous change in the slope of the  heavy fields potential). Since the potential of the light field is untouched and background quantities are continuous, the light field's perturbations are unaffected by the transition, i.e., $[Q_{\mathrm l}]^{\pm}=[\dot{Q}_{\mathrm l}]^{\pm}=0$ in the flat gauge. 

Perturbations of the heavy field are sensitive to the change in its potential, but since $Q_{\mathrm l}$ as well as $\mathcal R$ and $\dot{\phi}_i$ are continuous, $Q_{\mathrm h}$ has to be continuous as well according to (\ref{eq:curvature}). 
Thus, only $\dot{Q_{\mathrm h}}$ obeys a non-trivial matching condition,
\begin{eqnarray}
   & \left[Q_{\mathrm l}\right]^\pm & =  0 \quad \, , \quad \left[Q_{\mathrm h}\right]^\pm = 0 \, , 
	 \label{match1}
	 \\
	 & \left[\dot{Q}_{\mathrm l}\right]^\pm & = 0 \quad \, , \quad \left[\dot{\phi}_{\mathrm h}\dot{Q}_{\mathrm h}+Q_{\mathrm h} \left(\partial_{\phi_{\mathrm h}}V+3H\dot{\phi}_{\mathrm h}\right)\right]^\pm = 0 
	 \label{match2}\,.
\end{eqnarray}

The Sasaki-Mukhanov variables in (\ref{uv}) with (\ref{solv-}) and (\ref{solv+}) before and after the transition at $t_*$ read
\begin{eqnarray}
   Q_{\mathrm h}^\pm  & = & \cos \theta_\pm \frac{v_{\mathrm h}^\pm}{a} - \sin\theta_\pm \frac{v_{\mathrm l}^\pm}{a} \, , 
	 \label{Qh+}
	 \\
	 Q_{\mathrm l}^\pm  & = & \sin \theta_\pm \frac{v_{\mathrm h}^\pm}{a} + \cos\theta_\pm \frac{v_{\mathrm l}^\pm}{a} \, .
	 \label{Ql+}
\end{eqnarray}

The solutions for $t>t_*$ in (\ref{solv+}) entail  Bogoliubov coefficients that have to be expanded in terms of Gaussian random fields $e_i(\vec{k})$ as
 \begin{eqnarray}
     A_{\mathrm l} & = & \alpha_{\mathrm l}e_{\mathrm l}(\vec{k})+\alpha_{\mathrm h}e_{\mathrm h}(\vec{k}) \quad \,, \quad A_{\mathrm h} = \tilde{\alpha}_{\mathrm l}e_{\mathrm l}(\vec{k})+\tilde{\alpha}_{\mathrm h}e_{\mathrm h}(\vec{k}) \,, \nonumber
		 \\
		 B_{\mathrm l} & = & \beta_{\mathrm l}e_{\mathrm l}(\vec{k})+\beta_{\mathrm h}e_{\mathrm h}(\vec{k}) \quad \,, \quad B_{\mathrm h} = \tilde{\beta}_{\mathrm l}e_{\mathrm l}(\vec{k})+\tilde{\beta}_{\mathrm h}e_{\mathrm h}(\vec{k}) \,.
		 \label{expBogo}
 \end{eqnarray}
 
 It is straightforward to compute these coefficients by using the solutions  in (\ref{Qh+}) and (\ref{Ql+}) in the matching conditions (\ref{match1}), see App.~\ref{App:1} for the lengthy, explicit expressions.

\begin{figure}[tb]
   \begin{center}
      \includegraphics[width=0.44\columnwidth]{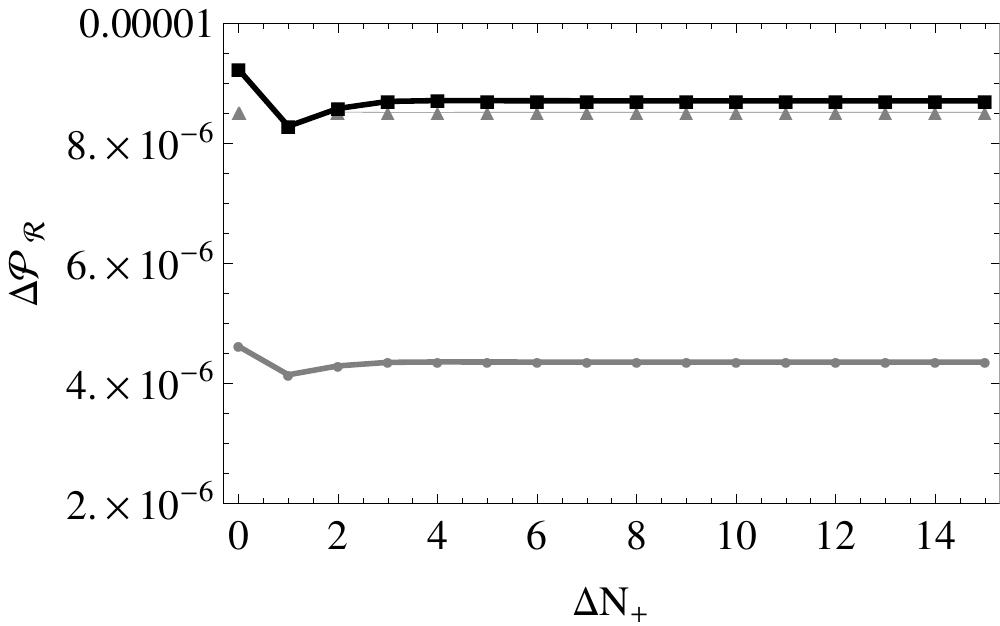}
			\includegraphics[width=0.55\columnwidth]{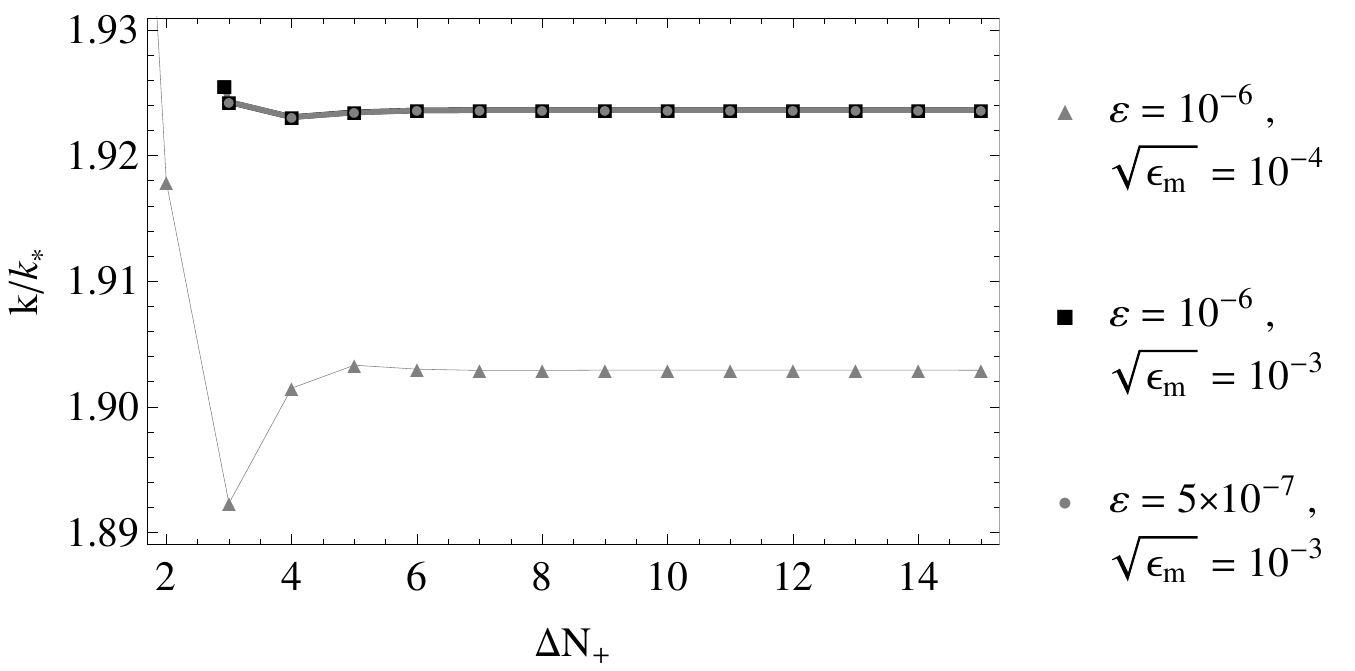}
	    \caption{The correction to the power-spectrum at the first peak and the position of the first peak as a function of the number of e-folds after the transition $\Delta N_+$ for various combinations of $\varepsilon$ and $\epsilon_{\mathrm m}$, with $\phi_{\mathrm l}^*=15 M_{\mathrm{Pl}}$. Waiting a few e-folds suffices to enter an adiabatic regime where (\ref{fullpowerspectrum}) can be used.\label{Fig:efolddependence}}
	 \end{center}		
\end{figure}

A few e-folds after the transition, the energy of the heavy field becomes negligible compared to the light one; subsequently, the light field decays and reheats the universe. As a consequence, its perturbations $Q_{\mathrm l}^+$ determine the fluctuations in the CMBR; at late times ($-k\tau \rightarrow 0$) the latter are therefore set by 
\begin{equation}
   \mathcal{R} \approx \frac{H}{\dot{\phi}_{\mathrm l}}Q_{\mathrm l} \,. 
\end{equation}
Further, we can approximate the Hankel functions in (\ref{Ql+}) by
\begin{equation}
   \mathcal{H}_{\mu_i^+}^{(2)} \approx - \mathcal{H}_{\mu_i^+}^{(1)} \approx \frac{\dot{\imath}}{\pi}\Gamma\left(\mu_i^+\right)\left(-\frac{k\tau_+}{2}\right)^{-\mu_i^+} \,,
\end{equation}
which, together with (\ref{expBogo}), leads to 
\begin{eqnarray}
   Q_{\mathrm l}^+ & \approx & -\dot{\imath}\frac{\sqrt{-\tau_+}}{a}\frac{\Gamma(\mu_{\mathrm l}^+)}{\sqrt{\pi}}\left(-\frac{k\tau_+}{2}\right)^{-\mu_{\mathrm l}^+}\label{QLfinal}\\
   &&\nonumber\times \Bigg{\{}\left[(\alpha_{\mathrm l}-\beta_{\mathrm l})\cos \theta_+ + (\tilde{\alpha}_{\mathrm l}-\tilde{\beta}_{\mathrm l})\frac{\Gamma(\mu_{\mathrm h}^+)}{\Gamma(\mu_{\mathrm l}^+)}\left(-\frac{k\tau_+}{2}\right)^{\mu_{\mathrm l}^+ - \mu_{\mathrm h}^+}\sin \theta_+\right]e_{\mathrm l}   \\ 
\nonumber	&& +  \left[(\alpha_{\mathrm h}-\beta_{\mathrm h})\cos \theta_+ + (\tilde{\alpha}_{\mathrm h}-\tilde{\beta}_{\mathrm h})\frac{\Gamma(\mu_{\mathrm h}^+)}{\Gamma(\mu_{\mathrm l}^+)}\left(-\frac{k\tau_+}{2}\right)^{\mu_{\mathrm l}^+ - \mu_{\mathrm h}^+}\sin \theta_+\right]e_{\mathrm h}\Bigg{\}} \,.
\end{eqnarray}
Here, the $\alpha_i, \beta_i$ with $i=l,h$ are given by (\ref{alphalfinal}), (\ref{betalfinal}), (\ref{alphahfinal}) and (\ref{betahfinal}) in the Appendix.

\begin{figure}[t]
   \begin{center}
      \includegraphics[width=0.70\columnwidth]{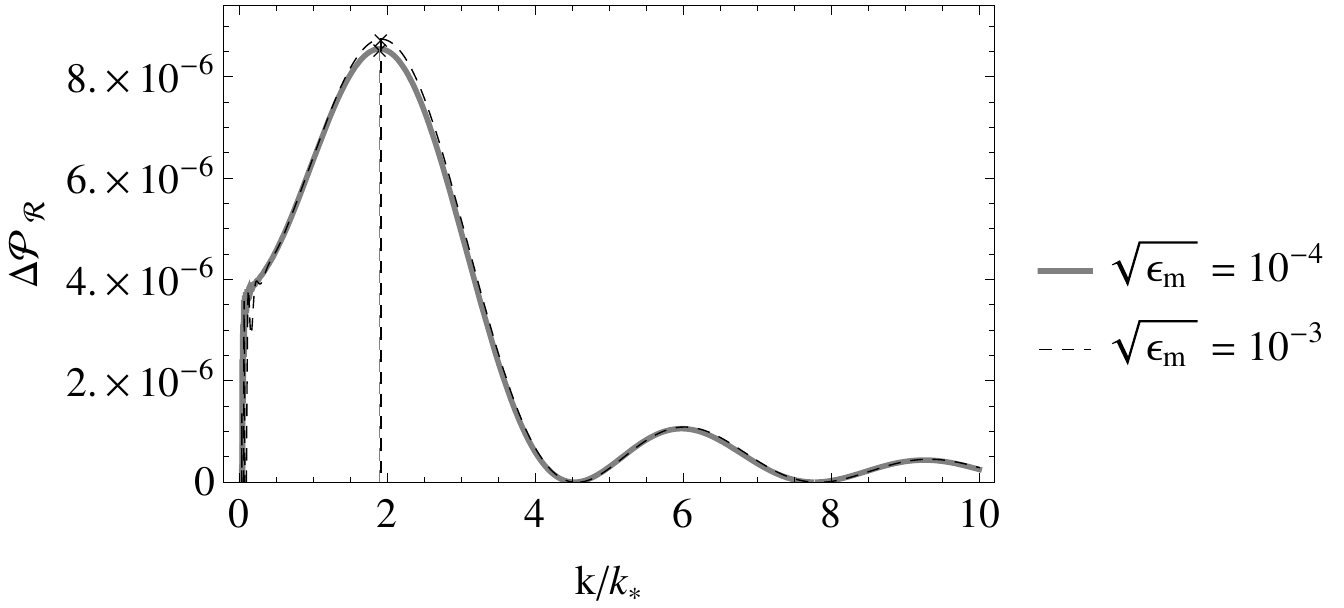}
	    \caption{The correction to the power-spectrum with $\Delta N_+=10$, $\varepsilon = 10^{-6}$ and $\phi_{\mathrm l}^*=15\, M_{\mathrm{Pl}}$. The first peak of the gray curve has an amplitude of $8.54 \times 10^{-6}$ at $k/k_* = 1.90$ ($\sqrt{\epsilon_{\mathrm m}} = 10^{-4}$) and the peak of the dashed curve has an amplitude of $8.74 \times 10^{-6}$ at $k/k_* = 1.92$ ($\sqrt{\epsilon_{\mathrm m}} =10 ^{-3}$). Evidently, only a weak dependence on $\epsilon_{\mathrm{m}}$ is present.}
	 \label{delta_p_3}
	 \end{center}		
\end{figure}

\subsubsection{The Power-spectrum} 
Given the solution in (\ref{QLfinal}), the power-spectrum reads
\begin{eqnarray}
   \mathcal{P}_{\mathcal{R}} & \approx & \left(\frac{H^2}{2\pi \dot{\phi}_{\mathrm l}}\frac{\Gamma(\mu_{\mathrm l}^+)}{\Gamma(3/2)}\right)^2\left|-\frac{k\tau_+}{2}\right|^{3-2\mu_{\mathrm l}^+}\label{fullpowerspectrum}\\
   \nonumber &&\times \Bigg{(}\left|(\alpha_{\mathrm l}-\beta_{\mathrm l})\cos \theta_+ + (\tilde{\alpha}_{\mathrm l}-\tilde{\beta}_{\mathrm l})\frac{\Gamma(\mu_{\mathrm h}^+)}{\Gamma(\mu_{\mathrm l}^+)}\left(-\frac{k\tau_+}{2}\right)^{\mu_{\mathrm l}^+ - \mu_{\mathrm h}^+}\sin \theta_+\right|^2 \nonumber \\ 
	& &+ \left|(\alpha_{\mathrm h}-\beta_{\mathrm h})\cos \theta_+ + (\tilde{\alpha}_{\mathrm h}-\tilde{\beta}_{\mathrm h})\frac{\Gamma(\mu_{\mathrm h}^+)}{\Gamma(\mu_{\mathrm l}^+)}\left(-\frac{k\tau_+}{2}\right)^{\mu_{\mathrm l}^+ - \mu_{\mathrm h}^+}\sin \theta_+\right|^2\Bigg{)} \, .
\end{eqnarray}

Evidently, the cumbersome analytic expressions for $\alpha_i$ and $\beta_i$ make this result somewhat opaque and less instructive than the one based on the sudden transition approximation in (\ref{powerspectrumcorrectionapprox}). We discuss the dependence of this oscillatory correction onto model parameter in the next section, where we also make contact with (\ref{powerspectrumcorrectionapprox}).

\subsubsection{Discussion \label{discussion2}}

We first note that the power-spectrum should only be computed once the system entered the adiabatic regime at a time when the energy density of the heavy field can safely be ignored and  the approximation in (\ref{fullpowerspectrum}) can be used. As $\rho_{\mathrm h}$ redshift quickly, it suffices to wait a few e-folds, see Fig.~\ref{delta_p_1} and \ref{Fig:efolddependence}, where we plot the height and position of the first peak of the full power-spectrum over the number of e-folds for some exemplary cases. In all subsequent plots we take $\Delta N_+=10$.

\begin{figure}[t]
   \begin{center}
      \includegraphics[width=0.44\columnwidth]{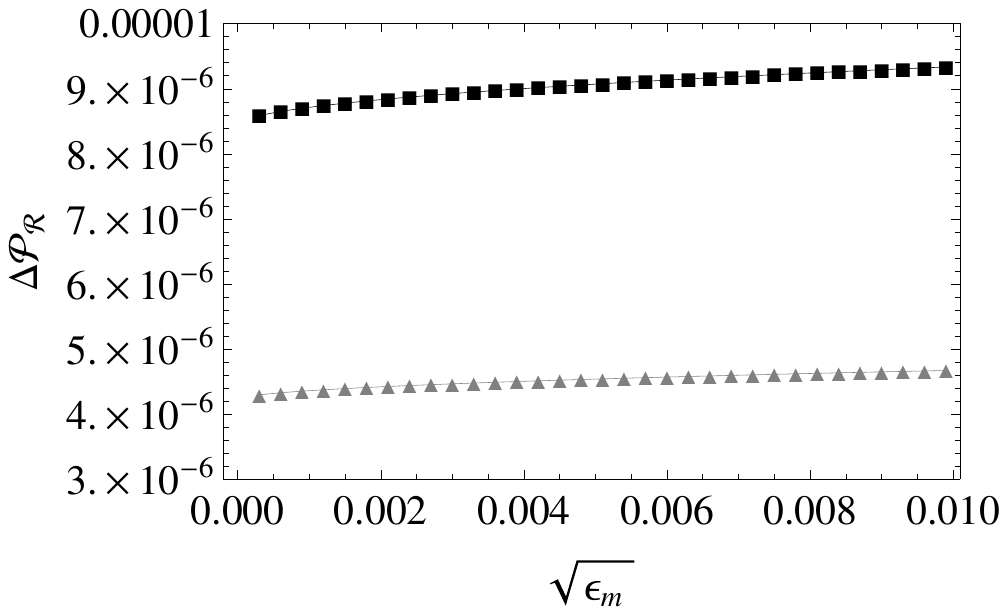}
			\includegraphics[width=0.53\columnwidth]{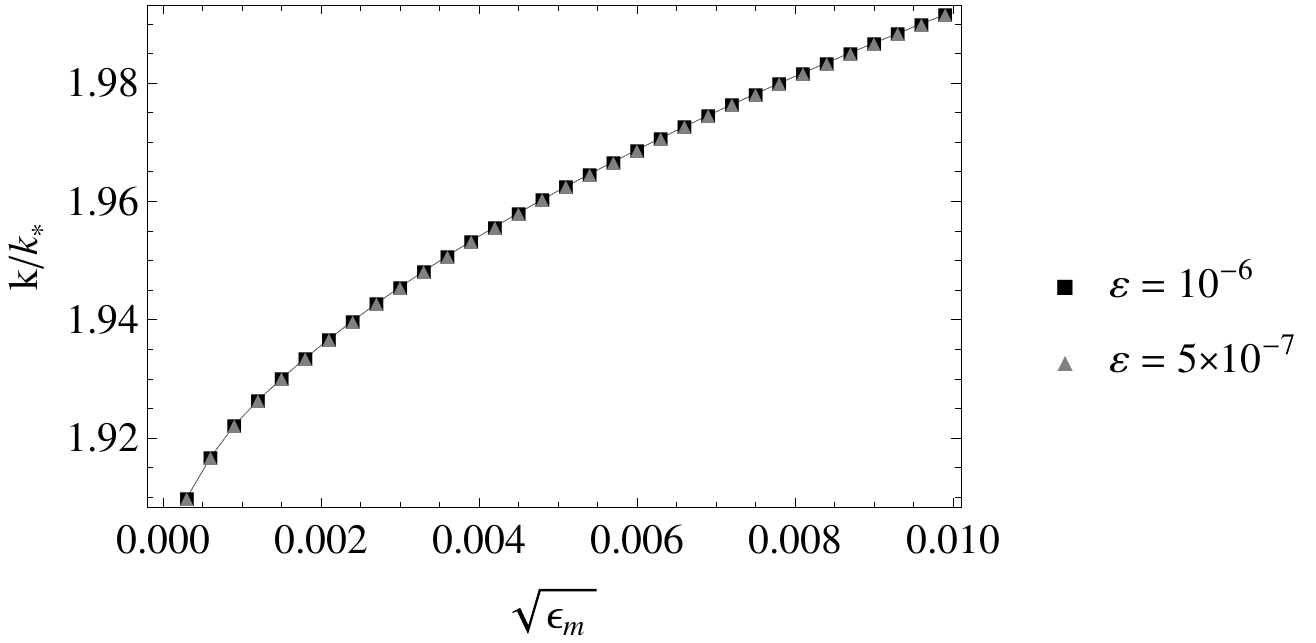}
	    \caption{The first peaks amplitude of the correction to the power-spectrum and its position as a function of  $\sqrt{\epsilon_{\mathrm m}}$ for two values of $\varepsilon$, with $\Delta N_+=10$ and $\phi_{\mathrm l}^*=15 M_{\mathrm{Pl}}$. Evidently, only a weak dependence on $\sqrt{\epsilon_{\mathrm{m}}}$ is present. \label{delta_p_3_2}}
	 \end{center}		
\end{figure}

According to the sudden transition approximation, the oscillatory correction should not depend on the ratio of the masses as parametrized by $\epsilon_m=m_{\mathrm l}^2/m_{\mathrm h}^2$. In Fig.~\ref{delta_p_3}, we observe that this is indeed a good approximation, since both, the position and height of the first peak change at the percent level only under a change of
$\epsilon_m$ by a factor of $100$. In Fig.~\ref{delta_p_3_2} we plot the height and position of the first peak for a range of $\sqrt{\epsilon_{\mathrm m}}$: it is clear that the dependence on $\sqrt{\epsilon_{\mathrm m}}$ is not even linear and thus negligible for the range of mass ratios that we are interested in ($\epsilon_{\mathrm m}\ll 1$).

\begin{figure}[t]
   \begin{center}
      \includegraphics[width=0.70\columnwidth]{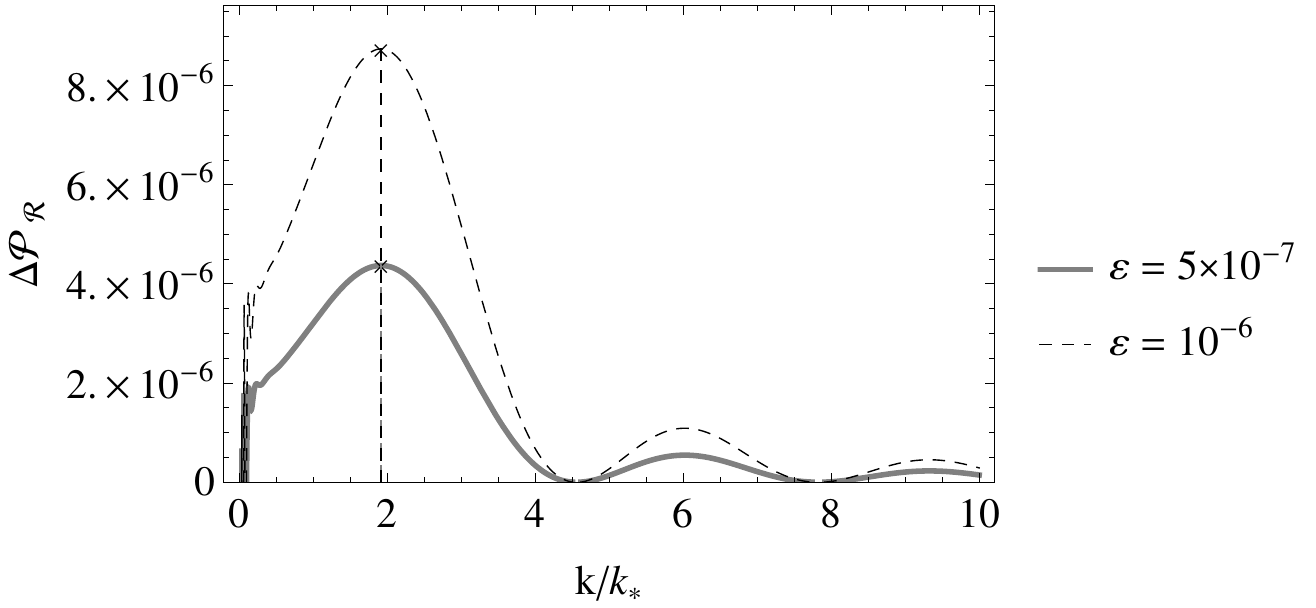}
	    \caption{The correction to the power-spectrum with $\Delta N_+=10$, $\sqrt{\epsilon_{\mathrm m}} = 10^{-3}$ and $\phi_{\mathrm l}^*=15 M_{\mathrm{Pl}}$. The first peak of both curves is located at $k/k_* = 1.92$ with an amplitude of $4.37 \times 10^{-6}$ for the the gray curve ($\varepsilon = 5\times 10^{-7}$) and $8.74 \times 10^{-6}$ for the the black dashed curve ($\varepsilon = 10^{-6}$). The linear dependence on $\varepsilon$ is in line with the sudden transition approximation in (\ref{powerspectrumcorrectionapprox}).}
	 \label{delta_p_2}
	 \end{center}		
\end{figure}

\begin{figure}[t]
   \begin{center}
      \includegraphics[width=0.44\columnwidth]{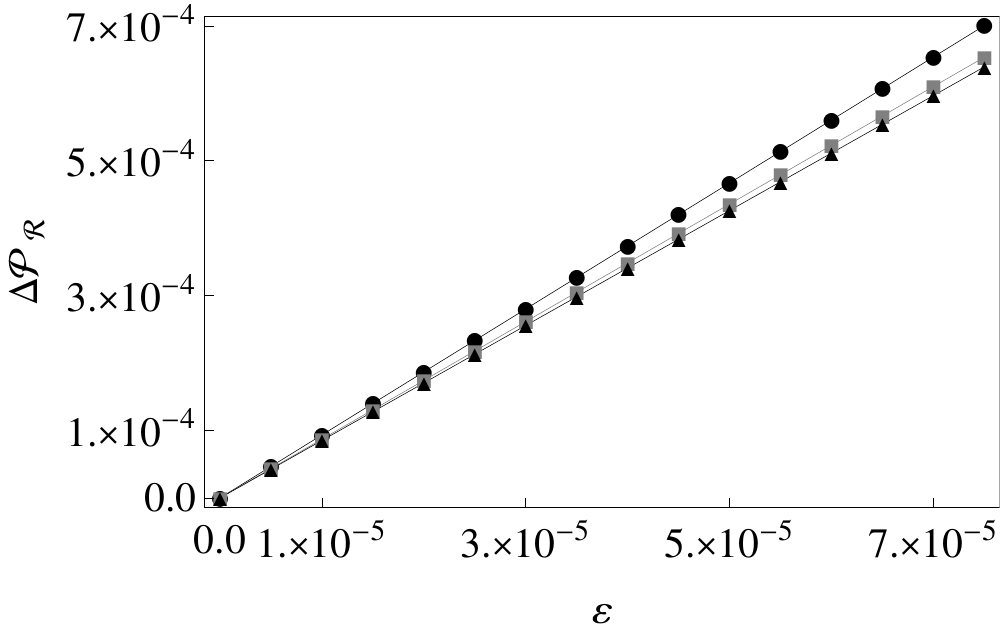}
			\includegraphics[width=0.53\columnwidth]{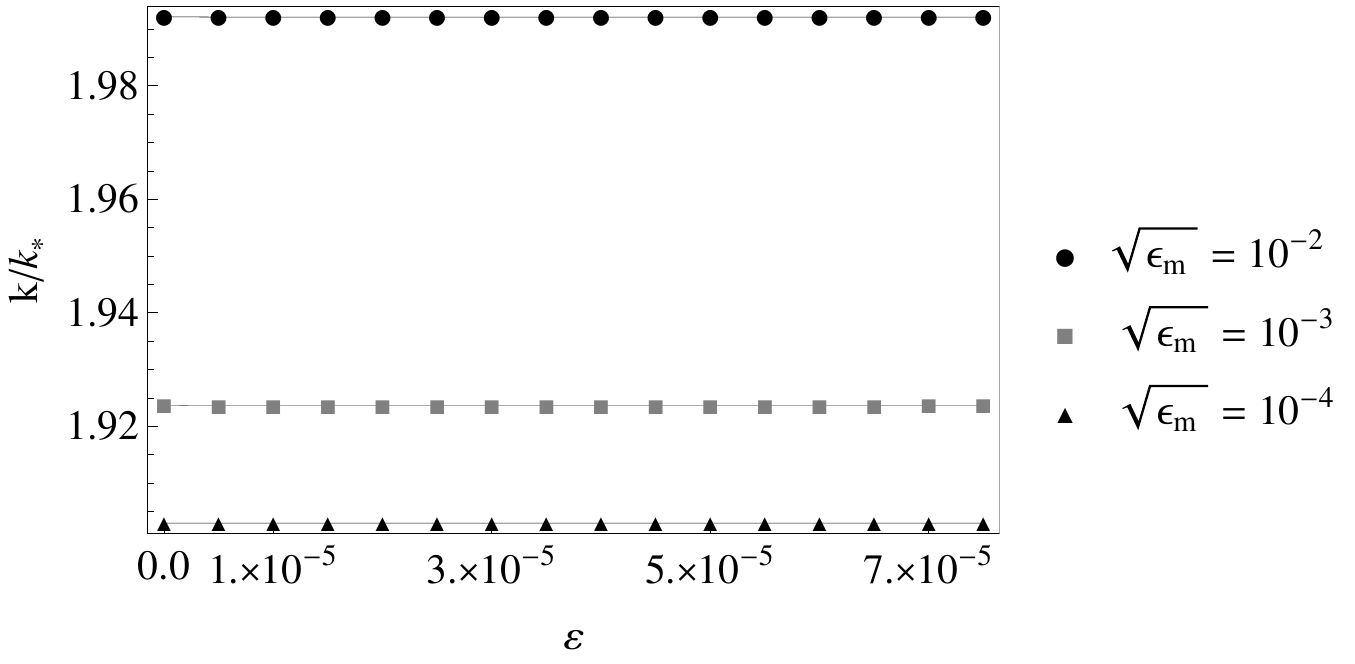}
	    \caption{The correction to the power-spectrum at the first peak and the position of the first peak as a function of  $\varepsilon$ for various values of $\epsilon_{\mathrm m}$ with  $\Delta N_+=10$ and $\phi_{\mathrm l}^*=15 M_{\mathrm{Pl}}$. In the left panel the inclinations of the curves are from top to bottom $9.34$, $8.72$ and $8.52$.  The linear dependence on $\varepsilon$ is in line with the sudden transition approximation in (\ref{powerspectrumcorrectionapprox}) although minor quantitative differences are present, such as the weak dependence of the slope on $\sqrt{\epsilon_{\mathrm{m}}}$.}
	 \label{k_epsilon}
	 \end{center}		
\end{figure}

The deciding factor determining the amplitude of the oscillatory correction is the energy ratio $\varepsilon=\rho_{\mathrm h}/\rho_{\mathrm l}$,  see Fig.~\ref{delta_p_2} and \ref{k_epsilon}. This result is expected based on the sudden transition approximation, see (\ref{powerspectrumcorrectionapprox}) and the discussion in Sec.~\ref{discussion1}. The dependence is linear, see Fig.~\ref{k_epsilon}, with a slope that is only weakly dependent on the mass ratio $\sqrt{\epsilon_{\mathrm m}}$.

\begin{figure}[t]
   \begin{center}
      \includegraphics[width=0.60\columnwidth]{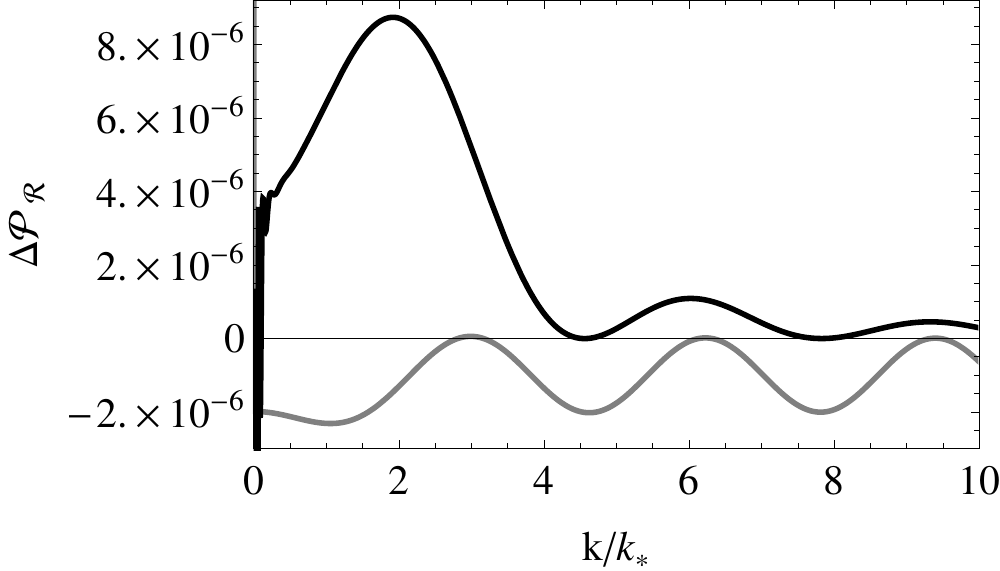}
	    \caption{A comparison between the sudden transition approximation (gray curve) and the full two field setup (black curve) with $\Delta N_+ = 10$, $\varepsilon = 10^{-6}$, $\sqrt{\epsilon_{\mathrm m}} = 10^{-3}$ and $\phi_{\mathrm l} = 15 M_{\mathrm{Pl}}$. The simplistic sudden transition approximation provides a good estimate for the amplitude and frequency of the correction to the power-spectrum. \label{Fig:comparison}}
	 \end{center}		
\end{figure}

Turning to a direct comparison of the sudden transition approximation in (\ref{powerspectrumcorrectionapprox}) with the two field result in (\ref{fullpowerspectrum}), we observe indeed a good qualitative agreement, see Fig.~\ref{Fig:comparison}; the main difference is a damping in the amplitude for increasing $k$ in the full result. This damping is caused by the finite duration of the transition, which is resolved in the full result, but absent in the sudden transition approximation. As a consequence, all modes, no matter how deep in the horizon, are affected in the sudden transition approximation (an artefact of the approximation). The amplitude (peak to valley) in (\ref{powerspectrumcorrectionapprox}) is given by $2\varepsilon$ and lies between the amplitude of the first and second peak of the full result in  (\ref{fullpowerspectrum}), see Fig.~\ref{Fig:comparison}. This is as good of a quantitative agreement as one could hope for. Fruther, the frequency is correctly recovered, but not the phase (that the phase matches reasonably well in Fig.~\ref{Fig:comparison} is a coincidence). Lastly, we observe a different sign of the correction in (\ref{powerspectrumcorrectionapprox}) and (\ref{fullpowerspectrum}) (positive for the full result and negative for the approximation).

We conclude that (\ref{fullpowerspectrum}) provides a good approximation to the full result as long as one keeps in mind that
\begin{itemize}
\item Oscillatory corrections are damped for increasing $k$ (an effect that can't be recovered in the sudden transition approximation).
\item The phase of the corrections can not be recovered, since the intermediate regime is cut out in the approximation.
\item The sign of the correction in the sudden transition approximation is opposite to the full result in the particular excitation mechanism considered in this article.
\end{itemize}

Thus, the maximal amplitude and frequency of the correction are recovered well. Since the peak's position is a free parameter (set by timing of the excitation event), the phase does not carry any physically relevant information if one is only concerned with the power-spectrum. However, if the signal is to be cross correlated with oscillations in the bi-spectrum as in \cite{Mizuno:2014jja,Achucarro:2012fd,Achucarro:2013cva,Gong:2014spa,Achucarro:2014msa}, the phase is important. Therefore, the approximation in (\ref{powerspectrumcorrectionapprox}) can provide a correct order of magnitude estimate, but for comparison with data more sophisticated techniques should be used.

We would like to reiterate that this estimate and the conclusions in Sec.~\ref{discussion1} apply to \emph{any} excitation event of a heavy field during inflation: they only depend on the change of the equation of state parameter well before and after the transition. This jump is present for \emph{any} excitation event, since $\rho_{\mathrm{h}}$ red-shifts by definition faster than the energy density driving inflation.

Additional, model dependent contributions are present if one goes beyond the sudden transition approximation; whenever these are computed, for instance within an EFT or via numerical techniques, the excitation of the heavy field still needs to satisfy the bound on $\varepsilon$ in (\ref{conditiononenergy}), which entails a bound on the heavy field's amplitude once a potential is specified, see equation (\ref{conditiononamplitude}).

The concrete setup discussed in this paper combined with the full two field treatment of perturbations in this section provides additional evidence supporting the universality of the estimate based on the sudden transition approximation. Thus, we advocate to use (\ref{conditiononenergy}) as a quick consistency check for any study dealing with excitations of heavy fields during inflation.

\section{Conclusion}

The aftermath of a heavy field's temporary excitation during inflation has been the focus of several recent studies, leading to the development of effective field theories with a time dependent speed of sound, approximate analytic schemes to recover signals caused by resonances and numerical studies. A common effect is the generation of oscillatory contributions to correlation functions, which are strongly constrained by CMBR observations. Since heavier fields are common in all inflationary models in string theory, such effects offer a window of opportunity onto dynamics that go beyond canonical, single field, slow-roll inflation.

In this paper, we investigated the effect of an excitation event  onto the power-spectrum, which necessarily entails a departure of slow-roll for the heavy field even if inflation is un-interrupted. Based on current bounds on oscillations in the power-spectrum ($\mathcal{O}(10^{-2})$ according to PLANCK), we arrived at a universal upper limit on the energy that can be infused into a heavy field during inflation
\begin{eqnarray}
\frac{\rho_{\mathrm{heavy}}}{\rho_{\mathrm{infl.}}}\lesssim \Delta{\mathcal{P_R}} \sim \mathcal{O}(10^{-2}) \,.\label{finalsummary}
\end{eqnarray}
 The oscillations leading to this bound can be traced back to the jump of the equation of state parameter from the value before the excitation to the one after the event, long after the temporary oscillations in the heavy field have died down. Hence, they are not due to any resonance effects or complicated interactions of perturbations during the intermediate regime, but a mere reflection of the fact that slow-roll dynamics changed abruptly  (without disturbing slow-roll of the light field). Thus, this bound applies to any excitation event, be it a sharp turn, a sudden mass change for one of the inflatons or other mechanisms.

We derived this bound in a sudden transition approximation, which cuts out the intermediate, non-trivial dynamics of the heavy field and truncates fluctuations to the light sector. This approximation is not new and the result is identical to the one derived for decaying inflatons in \cite{Battefeld:2010rf}. However, it is far from obvious that this simplistic approximation provides a good estimate: the temporary deviation of the equation of state parameter during the excitation event is considerably bigger than its asymptotic change relevant for the estimate, see Fig.~\ref{fig.EoS_param}; furthermore, fluctuations are expected to interact. Therefore, in a concrete two field setup where one field becomes heavy, we computed perturbations of both fields and kept the effect of the intermediate regime when applying the Deruelle-Mukhanov matching conditions, which need to be imposed once the mass of one field makes a jump. We found a good qualitative agreement to the much simpler result based on the sudden transition approximation. Further, similarly non-trivial studies of sharp turns, as in \cite{Konieczka:2014zja}, are also consistent with this estimate.

The reason why the latter approximation does a good job is twofold: firstly, slow-roll of the light field is not violated, so that background quantities, such as the Hubble parameter, experience only minor alterations; secondly, since the energy ratio $\rho_{\mathrm{heavy}}/\rho_{\mathrm{infl}}$  has to be less than one percent, the truncation to the light sector at the perturbed level becomes justified in retrospective (a necessary condition for the latter was derived in \cite{Battefeld:2013xka}).

We note that additional signals are caused by resonances brought forth by the subsequent oscillations of the heavy field (suppressed for canonical kinetic terms and therefore usually sub-leading) and interactions of fluctuations, both of which are not recovered. However, equation (\ref{finalsummary}) provides an excellent, simple rule of thumb: if too much energy is transferred to the heavy field, the excitation event is already ruled out and more complicated calculations are not needed. On the other hand, if, according to the estimate, a signal appears to have a desirable amplitude and frequency, it is prudent to go beyond the sudden transition approximation and include interactions as well as resonance effects; these are crucial for a meaningful comparison with observations, particularly if signals in the power-spectrum are to be cross correlated with the bi-spectrum.

\acknowledgments
We would like to thank Xingang Chen for comments. R. C. Freitas thanks the Institut f\"{u}r Astrophysik at the  Universit\"{a}t G\"{o}ttingen for the kind hospitality and the financial support by CAPES (Brazil), CNPq (Brazil) and DAAD (Germany). T.B.~would like to thank P.~Peter and the Institut Astrophysique de Paris (IAP) for hospitality during the final stages of this project.

\appendix
\section{Appendix \label{App:1}}
To calculate the explicit expressions for the Bogoliubov coefficients in (\ref{expBogo}) we need to plug the solutions in (\ref{Qh+}) and (\ref{Ql+}) into the matching conditions in (\ref{match1}), which yields
	\begin{eqnarray}
	   \alpha_{\mathrm l}\mathcal{H}_{\mu_{\mathrm l}^+}^{(1)}\cos \theta_+ + \beta_{\mathrm l}\mathcal{H}_{\mu_{\mathrm l}^+}^{(2)}\cos \theta_+ +\tilde{\alpha}_{\mathrm l}\mathcal{H}_{\mu_{\mathrm h}^+}^{(1)}\sin \theta_+ + \tilde{\beta}_{\mathrm l}\mathcal{H}_{\mu_{\mathrm h}^+}^{(2)}\sin \theta_+ & = & e^{\dot{\imath}\left(\mu_{\mathrm l}^-+1/2\right)\pi/2}\mathcal{H}_{\mu_{\mathrm l}^-}^{(1)}\cos \theta_- \, , \nonumber \\
		 \\
		\alpha_{\mathrm h}\mathcal{H}_{\mu_{\mathrm l}^+}^{(1)}\cos \theta_+ + \beta_{\mathrm h}\mathcal{H}_{\mu_{\mathrm l}^+}^{(2)}\cos \theta_+ +\tilde{\alpha}_{\mathrm h}\mathcal{H}_{\mu_{\mathrm h}^+}^{(1)}\sin \theta_+ + \tilde{\beta}_{\mathrm h}\mathcal{H}_{\mu_{\mathrm h}^+}^{(2)}\sin \theta_+ & = & e^{\dot{\imath}\left(\mu_{\mathrm h}^-+1/2\right)\pi/2}\mathcal{H}_{\mu_{\mathrm h}^-}^{(1)}\sin \theta_- \, , \nonumber \\
		 \\
		\alpha_{\mathrm l}\mathcal{H}_{\mu_{\mathrm l}^+}^{(1)}\sin \theta_+ + \beta_{\mathrm l}\mathcal{H}_{\mu_{\mathrm l}^+}^{(2)}\sin \theta_+ -\tilde{\alpha}_{\mathrm l}\mathcal{H}_{\mu_{\mathrm h}^+}^{(1)}\cos \theta_+ - \tilde{\beta}_{\mathrm l}\mathcal{H}_{\mu_{\mathrm h}^+}^{(2)}\cos \theta_+ & = & e^{\dot{\imath}\left(\mu_{\mathrm l}^-+1/2\right)\pi/2}\mathcal{H}_{\mu_{\mathrm l}^-}^{(1)}\sin \theta_- \, , \nonumber \\
		 \\
		\alpha_{\mathrm h}\mathcal{H}_{\mu_{\mathrm l}^+}^{(1)}\sin \theta_+ + \beta_{\mathrm h}\mathcal{H}_{\mu_{\mathrm l}^+}^{(2)}\sin \theta_+ -\tilde{\alpha}_{\mathrm h}\mathcal{H}_{\mu_{\mathrm h}^+}^{(1)}\cos \theta_+ - \tilde{\beta}_{\mathrm h}\mathcal{H}_{\mu_{\mathrm h}^+}^{(2)}\cos \theta_+ & = & -e^{\dot{\imath}\left(\mu_{\mathrm h}^-+1/2\right)\pi/2}\mathcal{H}_{\mu_{\mathrm h}^-}^{(1)}\cos \theta_- \,.\nonumber \\
	\end{eqnarray}
Furthermore, the matching conditions in (\ref{match2}) require
\begin{eqnarray}
  & & \alpha_{\mathrm l}\left(\frac{3}{2}\mathcal{H}_{\mu_{\mathrm l}^+}^{(1)}+\frac{k}{aH}\frac{d\mathcal{H}_{\mu_{\mathrm l}^+}^{(1)}}{dx}\right)\cos \theta_+ +\beta_{\mathrm l}\left(\frac{3}{2}\mathcal{H}_{\mu_{\mathrm l}^+}^{(2)}+\frac{k}{aH}\frac{d \mathcal{H}_{\mu_{\mathrm l}^+}^{(2)}}{dx}\right)\cos \theta_+  \nonumber \\
	&  & +\tilde{\alpha}_{\mathrm l}\left(\frac{3}{2}\mathcal{H}_{\mu_{\mathrm h}^+}^{(1)}+\frac{k}{aH}\frac{d\mathcal{H}_{\mu_{\mathrm h}^+}^{(1)}}{dx}\right)\sin \theta_+ +\tilde{\beta}_{\mathrm l}\left(\frac{3}{2}\mathcal{H}_{\mu_{\mathrm h}^+}^{(2)}+\frac{k}{aH}\frac{d\mathcal{H}_{\mu_{\mathrm h}^+}^{(2)}}{dx}\right)\sin \theta_+ \nonumber \\
	& = & e^{\dot{\imath}\left(\mu_{\mathrm l}^-+1/2\right)\pi/2}\left(\frac{3}{2}\mathcal{H}_{\mu_{\mathrm l}^-}^{(1)}+\frac{k}{aH}\frac{d\mathcal{H}_{\mu_{\mathrm l}^-}^{(1)}}{dx}\right)\cos \theta_- \, ,
\\
	& & \alpha_{\mathrm h}\left(\frac{3}{2}\mathcal{H}_{\mu_{\mathrm l}^+}^{(1)}+\frac{k}{aH}\frac{d\mathcal{H}_{\mu_{\mathrm l}^+}^{(1)}}{dx}\right)\cos \theta_+ +\beta_{\mathrm h}\left(\frac{3}{2}\mathcal{H}_{\mu_{\mathrm l}^+}^{(2)}+\frac{k}{aH}\frac{d\mathcal{H}_{\mu_{\mathrm l}^+}^{(2)}}{dx}\right)\cos \theta_+  \nonumber \\
	& &+ \tilde{\alpha}_{\mathrm h}\left(\frac{3}{2}\mathcal{H}_{\mu_{\mathrm h}^+}^{(1)}+\frac{k}{aH}\frac{d\mathcal{H}_{\mu_{\mathrm h}^+}^{(1)}}{dx}\right)\sin \theta_+ +\tilde{\beta}_{\mathrm h}\left(\frac{3}{2}\mathcal{H}_{\mu_{\mathrm h}^+}^{(2)}+\frac{k}{aH}\frac{d\mathcal{H}_{\mu_{\mathrm h}^+}^{(2)}}{dx}\right)\sin \theta_+ \nonumber \\
	& = & e^{\dot{\imath}\left(\mu_{\mathrm h}^-+1/2\right)\pi/2}\left(\frac{3}{2}\mathcal{H}_{\mu_{\mathrm h}^-}^{(1)}+\frac{k}{aH}\frac{d\mathcal{H}_{\mu_{\mathrm h}^-}^{(1)}}{dx}\right)\sin \theta_- \,,
\end{eqnarray}
\begin{table}
	 \begin{center}
   \begin{tabular}[t]{|c | c c | c c || c c|}
       \hline
       ~ & $\cot \gamma$ &  & $\gamma \approx \pi/2$ & ~ & Diff. ($\%$) & ~ \\ \cline{2-7}
       ~ & $\delta \mathcal{P}_{\mathcal{R}}^*$ & $k/k_*$ & $\delta \mathcal{P}_{\mathcal{R}}^*$ & $k/k_*$ & $\delta \mathcal{P}_{\mathcal{R}}^*$ & $k/k_*$ \\ \hline
		   Figure 4g & $8.66 \times 10^{-6}$ & $1.92$ & $ 8.33 \times 10^{-6}$ & $2.00$ & $-3.81$ & $4.17$\\
		   Figure 4b & $8.74 \times 10^{-6}$ & $1.92$ & $ 8.28 \times 10^{-6}$ & $1.95$ & $-5.26$ & $1.56$\\
		   Figure 5g & $4.37 \times 10^{-6}$ & $1.92$ & $ 4.14 \times 10^{-6}$ & $1.95$ & $-5.26$ & $1.56$\\
		   Figure 5b & $8.74 \times 10^{-6}$ & $1.92$ & $ 8.28 \times 10^{-6}$ & $1.95$ & $-5.26$ & $1.54$\\
		   Figure 6g & $8.54 \times 10^{-6}$ & $1.90$ & $ 8.08 \times 10^{-6}$ & $1.93$ & $-5.39$ & $1.58$\\
		   Figure 6b & $8.74 \times 10^{-6}$ & $1.92$ & $ 8.28 \times 10^{-6}$ & $1.95$ & $-5.26$ & $1.56$\\ \hline
   \end{tabular}
	 \end{center}
	 \caption{Comparision of the first peak amplitude and position for the case $\cot \gamma$ (that can be seen in Figures \ref{delta_p_1}--\ref{delta_p_3}) and for the case $\gamma \approx \pi/2$. Here ``g'' stands for the gray curve of the plots and ``b'' for the black curve.}
   \label{table1}
\end{table}
and
\begin{eqnarray}
&&\nonumber - e^{\dot{\imath}\left(\mu_{\mathrm l}^-+1/2\right)\pi/2}\left(\frac{3}{2}\mathcal{H}_{\mu_{\mathrm l}^-}^{(1)}+\frac{k}{aH}\frac{d\mathcal{H}_{\mu_{\mathrm l}^-}^{(1)}}{dx}\right)\sin \theta_- =\\
&&
 \alpha_{\mathrm l}\left[\frac{3}{2}\left(1+2\frac{V_{\mathrm h}^\prime}{3H\dot{\phi}_{\mathrm h}}\right)\mathcal{H}_{\mu_{\mathrm l}^+}^{(1)}+\frac{k}{aH}\frac{d\mathcal{H}_{\mu_{\mathrm l}^+}^{(1)}}{dx}\right]\sin \theta_+ \nonumber \\
	  &&+\beta_{\mathrm l}\left[\frac{3}{2}\left(1+2\frac{V_{\mathrm h}^\prime}{3H\dot{\phi}_{\mathrm h}}\right)\mathcal{H}_{\mu_{\mathrm l}^+}^{(2)}+\frac{k}{aH}\frac{d\mathcal{H}_{\mu_{\mathrm l}^+}^{(2)}}{dx}\right]\sin \theta_+ \nonumber \\
	& &- \tilde{\alpha}_{\mathrm l}\left[\frac{3}{2}\left(1+2\frac{V_{\mathrm h}^\prime}{3H\dot{\phi}_{\mathrm h}}\right)\mathcal{H}_{\mu_{\mathrm h}^+}^{(1)}+\frac{k}{aH}\frac{d\mathcal{H}_{\mu_{\mathrm h}^+}^{(1)}}{dx}\right]\cos \theta_+ 
\nonumber \\	
&&	-\tilde{\beta}_{\mathrm l}\left[\frac{3}{2}\left(1+2\frac{V_{\mathrm h}^\prime}{3H\dot{\phi}_{\mathrm h}}\right)\mathcal{H}_{\mu_{\mathrm h}^+}^{(2)}+\frac{k}{aH}\frac{d\mathcal{H}_{\mu_{\mathrm h}^+}^{(2)}}{dx}\right]\cos \theta_+  \,,\\	
\nonumber && e^{\dot{\imath}\left(\mu_{\mathrm h}^-+1/2\right)\pi/2}\left(\frac{3}{2}\mathcal{H}_{\mu_{\mathrm h}^-}^{(1)}+\frac{k}{aH}\frac{d\mathcal{H}_{\mu_{\mathrm h}^-}^{(1)}}{dx}\right)\cos \theta_- =\\
\nonumber &&
	 \alpha_{\mathrm h}\left[\frac{3}{2}\left(1+2\frac{V_{\mathrm h}^\prime}{3H\dot{\phi}_{\mathrm h}}\right)\mathcal{H}_{\mu_{\mathrm l}^+}^{(1)}+\frac{k}{aH}\frac{d\mathcal{H}_{\mu_{\mathrm l}^+}^{(1)}}{dx}\right]\sin \theta_+ \\
\nonumber &&	 +\beta_{\mathrm h}\left[\frac{3}{2}\left(1+2\frac{V_{\mathrm h}^\prime}{3H\dot{\phi}_{\mathrm h}}\right)\mathcal{H}_{\mu_{\mathrm l}^+}^{(2)}+\frac{k}{aH}\frac{d\mathcal{H}_{\mu_{\mathrm l}^+}^{(2)}}{dx}\right]\sin \theta_+ + \nonumber \\
	&  &- \tilde{\alpha}_{\mathrm h}\left[\frac{3}{2}\left(1+2\frac{V_{\mathrm h}^\prime}{3H\dot{\phi}_{\mathrm h}}\right)\mathcal{H}_{\mu_{\mathrm h}^+}^{(1)}+\frac{k}{aH}\frac{d\mathcal{H}_{\mu_{\mathrm h}^+}^{(1)}}{dx}\right]\cos \theta_+ \nonumber \\
&&	-\tilde{\beta}_{\mathrm h}\left[\frac{3}{2}\left(1+2\frac{V_{\mathrm h}^\prime}{3H\dot{\phi}_{\mathrm h}}\right)\mathcal{H}_{\mu_{\mathrm h}^+}^{(2)}+\frac{k}{aH}\frac{d\mathcal{H}_{\mu_{\mathrm h}^+}^{(2)}}{dx}\right]\cos \theta_+ 	
\end{eqnarray}

\begin{table}
   \begin{center}
   \begin{tabular}[t]{|c | c | c || c|}
       \hline
           ~     & $\cot \gamma$ & $\gamma \approx \pi/2$ & Diff. ($\%$) \\ \hline
           ~     & $9.34$ & $8.92$ & $-4.50$ \\
		    Figure 8  & $8.72$ & $8.26$ & $-5.28$ \\
		       ~     & $8.52$ & $8.06$ & $-5.40$ \\ \hline
   \end{tabular}
   \caption{Comparision of the curve slope for the case $\cot \gamma$ (that can be seen in Figure \ref{k_epsilon}) and for the case $\gamma \approx \pi/2$.}
   \label{table2}
\end{center}
\end{table}

Solving for the Bogoliubov coefficients, one arrives at
\begin{eqnarray}
   \alpha_{\mathrm l} & = & - \frac{\pi}{4}\dot{\imath}e^{\dot{\imath}\left(\mu_{\mathrm l}^-+1/2\right)\pi/2}\Bigg{\{}\cos \theta_- \cos \theta_+ (1+\tan \theta_- \tan \theta_+)\frac{k}{aH}\left[\mathcal{H}_{\mu_{\mathrm l}^+}^{(2)}\frac{d\mathcal{H}_{\mu_{\mathrm l}^-}^{(1)}}{dx}-\mathcal{H}_{\mu_{\mathrm l}^-}^{(1)}\frac{d\mathcal{H}_{\mu_{\mathrm l}^+}^{(2)}}{dx}\right]+ \nonumber \\
	& + & 3 \left(1+\frac{V_{\mathrm h}^\prime}{3H\dot{\phi}_{\mathrm h}}\right)\sin \theta_+\sin \theta_-\mathcal{H}_{\mu_{\mathrm l}^+}^{(2)}\mathcal{H}_{\mu_{\mathrm l}^-}^{(1)}\Bigg{\}} \, , \label{alphalfinal}	\\
 \beta_{\mathrm l} & = &  \frac{\pi}{4}\dot{\imath}e^{\dot{\imath}\left(\mu_{\mathrm l}^-+1/2\right)\pi/2}\Bigg{\{}\cos \theta_- \cos \theta_+ (1+\tan \theta_- \tan \theta_+)\frac{k}{aH}\left[\mathcal{H}_{\mu_{\mathrm l}^+}^{(1)}\frac{d\mathcal{H}_{\mu_{\mathrm l}^-}^{(1)}}{dx}-\mathcal{H}_{\mu_{\mathrm l}^-}^{(1)}\frac{d\mathcal{H}_{\mu_{\mathrm l}^+}^{(1)}}{dx}\right]+ \nonumber \\
	& + & 3\left(1+ \frac{V_{\mathrm h}^\prime}{3H\dot{\phi}_{\mathrm h}}\right)\sin \theta_-\sin \theta_+\mathcal{H}_{\mu_{\mathrm l}^+}^{(1)}\mathcal{H}_{\mu_{\mathrm l}^-}^{(1)}\Bigg{\}} \, , \label{betalfinal}\\
	 \tilde{\alpha}_{\mathrm l} & = & - \frac{\pi}{4}\dot{\imath}e^{\dot{\imath}\left(\mu_{\mathrm l}^-+1/2\right)\pi/2}\Bigg{\{}\cos \theta_- \sin \theta_+ \left(1-\frac{\tan \theta_-}{\tan \theta_+} \right)\frac{k}{aH}\left[\mathcal{H}_{\mu_{\mathrm h}^+}^{(2)}\frac{d\mathcal{H}_{\mu_{\mathrm l}^-}^{(1)}}{dx}-\mathcal{H}_{\mu_{\mathrm l}^-}^{(1)}\frac{d\mathcal{H}_{\mu_{\mathrm h}^+}^{(2)}}{dx}\right]+ \nonumber \\
	& - & 3 \left(1+\frac{V_{\mathrm h}^\prime}{3H\dot{\phi}_{\mathrm h}}\right)\sin \theta_-\cos \theta_+\mathcal{H}_{\mu_{\mathrm l}^-}^{(1)}\mathcal{H}_{\mu_{\mathrm h}^+}^{(2)}\Bigg{\}} \, , 
	\\
	\tilde{\beta}_{\mathrm l} & = & \frac{\pi}{4}\dot{\imath}e^{\dot{\imath}\left(\mu_{\mathrm l}^-+1/2\right)\pi/2}\Bigg{\{}\cos \theta_- \sin \theta_+ \left(1-\frac{\tan \theta_-}{\tan \theta_+} \right)\frac{k}{aH}\left[\mathcal{H}_{\mu_{\mathrm h}^+}^{(1)}\frac{d\mathcal{H}_{\mu_{\mathrm l}^-}^{(1)}}{dx}-\mathcal{H}_{\mu_{\mathrm l}^-}^{(1)}\frac{d\mathcal{H}_{\mu_{\mathrm h}^+}^{(1)}}{dx}\right]+ \nonumber \\
	& - & 3\left(1+\frac{V_{\mathrm h}^\prime}{3H\dot{\phi}_{\mathrm h}}\right)\sin \theta_-\cos \theta_+\mathcal{H}_{\mu_{\mathrm l}^-}^{(1)}\mathcal{H}_{\mu_{\mathrm h}^+}^{(1)}\Bigg{\}} \, , 
\\
	\alpha_{\mathrm h} & = & - \frac{\pi}{4}\dot{\imath}e^{\dot{\imath}\left(\mu_{\mathrm h}^-+1/2\right)\pi/2}\cos \theta_- \sin \theta_+\Bigg{\{}\left(1-\frac{\tan \theta_-}{\tan \theta_+}\right)\frac{k}{aH}\left[\mathcal{H}_{\mu_{\mathrm h}^-}^{(1)}\frac{d\mathcal{H}_{\mu_{\mathrm l}^+}^{(2)}}{dx}-\mathcal{H}_{\mu_{\mathrm l}^+}^{(2)}\frac{d\mathcal{H}_{\mu_{\mathrm h}^-}^{(1)}}{dx}\right]+ \nonumber \\
	& - & 3\left(1+ \frac{V_{\mathrm h}^\prime}{3H\dot{\phi}_{\mathrm h}}\right)\mathcal{H}_{\mu_{\mathrm l}^+}^{(2)}\mathcal{H}_{\mu_{\mathrm h}^-}^{(1)}\Bigg{\}} \, , \label{alphahfinal}
\\
	\beta_{\mathrm h} & = & \frac{\pi}{4}\dot{\imath}e^{\dot{\imath}\left(\mu_{\mathrm h}^-+1/2\right)\pi/2}\cos \theta_- \sin \theta_+\Bigg{\{}\left(1-\frac{\tan \theta_-}{\tan \theta_+}\right)\frac{k}{aH}\left[\mathcal{H}_{\mu_{\mathrm h}^-}^{(1)}\frac{d\mathcal{H}_{\mu_{\mathrm l}^+}^{(1)}}{dx}-\mathcal{H}_{\mu_{\mathrm l}^+}^{(1)}\frac{d\mathcal{H}_{\mu_{\mathrm h}^-}^{(1)}}{dx}\right]+ \nonumber \\
	&  &- 3\left(1+ \frac{V_{\mathrm h}^\prime}{3H\dot{\phi}_{\mathrm h}}\right)\mathcal{H}_{\mu_{\mathrm l}^+}^{(1)}\mathcal{H}_{\mu_{\mathrm h}^-}^{(1)}\Bigg{\}} \, , \label{betahfinal}
	\end{eqnarray}
\begin{eqnarray}
	\tilde{\alpha}_{\mathrm h} & = & - \frac{\pi}{4}\dot{\imath}e^{\dot{\imath}\left(\mu_{\mathrm h}^-+1/2\right)\pi/2}\cos \theta_- \cos \theta_+\Bigg{\{}\left(1+\tan \theta_-\tan \theta_+\right)\frac{k}{aH}\left[\mathcal{H}_{\mu_{\mathrm h}^+}^{(2)}\frac{d\mathcal{H}_{\mu_{\mathrm h}^-}^{(1)}}{dx}-\mathcal{H}_{\mu_{\mathrm h}^-}^{(1)}\frac{d\mathcal{H}_{\mu_{\mathrm h}^+}^{(2)}}{dx}\right]+ \nonumber \\
	& + & 3 \left(1+\frac{V_{\mathrm h}^\prime}{3H\dot{\phi}_{\mathrm h}}\right)\mathcal{H}_{\mu_{\mathrm h}^+}^{(2)}\mathcal{H}_{\mu_{\mathrm h}^-}^{(1)}\Bigg{\}} \, , 
	\\
	\tilde{\beta}_{\mathrm h} & = & \frac{\pi}{4}\dot{\imath}e^{\dot{\imath}\left(\mu_{\mathrm h}^-+1/2\right)\pi/2}\cos \theta_- \cos \theta_+\Bigg{\{}\left(1+\tan \theta_- \tan \theta_+\right)\frac{k}{aH}\left[\mathcal{H}_{\mu_{\mathrm h}^+}^{(1)}\frac{d\mathcal{H}_{\mu_{\mathrm h}^-}^{(1)}}{dx}-\mathcal{H}_{\mu_{\mathrm h}^-}^{(1)}\frac{d\mathcal{H}_{\mu_{\mathrm h}^+}^{(1)}}{dx}\right]+ \nonumber \\
	& + & 3\left(1+ \frac{V_{\mathrm h}^\prime}{3H\dot{\phi}_{\mathrm h}}\right)\mathcal{H}_{\mu_{\mathrm h}^+}^{(1)}\mathcal{H}_{\mu_{\mathrm h}^-}^{(1)}\Bigg{\}} \, ,
\end{eqnarray}
where all time dependent quantities are to be evaluated at $t_*$; at the transition we have $x\approx -k\tau_*$ and both, the angles $\theta_\pm$ as well as the Hankel function indices $\mu_i^{\pm}$, have to be evaluated at $ k_*\equiv a_*H_*$. 

We can approximate $\gamma \approx \pi/2$ in all expressions except in the ratio $V_{\mathrm h}^\prime/(3H\dot{\phi}_{\mathrm h})$: due to the condition $\epsilon_{\mathrm m} \ll 3\epsilon_{\mathrm l}$, the ratio $\epsilon_{\mathrm m}/(3\epsilon_{\mathrm l})$ in (\ref{gamma}) is second order when compared to $\epsilon_{\mathrm l}$;  if we were to use the approximation $\gamma \approx \pi/2$ in
\begin{equation}
   \frac{V_{\mathrm h}^{\prime}}{3H\dot{\phi}_{\mathrm h}}\Bigg{|}_{t=t_*} = 2\sqrt{\frac{\epsilon_{\mathrm l}}{3\epsilon_{\mathrm m}}}\left(\frac{1}{2\cot \gamma - \sqrt{3\epsilon_{\mathrm m}/\epsilon_{\mathrm l}}}\right) \,,
\end{equation}
we would ignore the leading order term, while keeping second order ones, which cancel if the full $\gamma$ from (\ref{gamma}) is used instead. Therefore we use the full expression in (\ref{gamma}) in all plots (the relative errors that result if the approximation $\gamma \approx \pi/2$ were used everywhere are given in Tab.~\ref{table1} and \ref{table2}).


\begin{thebibliography}{99}
\bibitem{Ade:2013uln} 
  P.~A.~R.~Ade {\it et al.}  [Planck Collaboration],
  arXiv:1303.5082 [astro-ph.CO].



\bibitem{Ade:2013ydc} 
  P.~A.~R.~Ade {\it et al.}  [Planck Collaboration],
  arXiv:1303.5084 [astro-ph.CO].



\bibitem{Ade:2013zuv} 
  P.~A.~R.~Ade {\it et al.}  [Planck Collaboration],
  arXiv:1303.5076 [astro-ph.CO].



\bibitem{Ade:2014xna} 
  P.~A.~R.~Ade {\it et al.}  [BICEP2 Collaboration],
  arXiv:1403.3985 [astro-ph.CO].



\bibitem{Bassett:2005xm} 
  B.~A.~Bassett, S.~Tsujikawa and D.~Wands,
  Rev.\ Mod.\ Phys.\  {\bf 78}, 537 (2006)
  [astro-ph/0507632].



\bibitem{Baumann:2009ds} 
  D.~Baumann,
  arXiv:0907.5424 [hep-th].



\bibitem{Martin:2013tda} 
  J.~Martin, C.~Ringeval and V.~Vennin,
  arXiv:1303.3787 [astro-ph.CO].



\bibitem{Baumann:2014nda} 
  D.~Baumann and L.~McAllister,
  arXiv:1404.2601 [hep-th].



\bibitem{Assassi:2013gxa} 
  V.~Assassi, D.~Baumann, D.~Green and L.~McAllister,
  arXiv:1304.5226 [hep-th].



\bibitem{Burgess:2012dz} 
  C.~P.~Burgess, M.~W.~Horbatsch and S.~.P.~Patil,
  JHEP {\bf 1301}, 133 (2013)
  [arXiv:1209.5701 [hep-th]].



\bibitem{Cespedes:2013rda} 
  S.~Céspedes and G.~A.~Palma,
  JCAP {\bf 1310}, 051 (2013)
  [arXiv:1303.4703 [hep-th]].



\bibitem{Avgoustidis:2012yc} 
  A.~Avgoustidis, S.~Cremonini, A.~-C.~Davis, R.~H.~Ribeiro, K.~Turzynski and S.~Watson,
  JCAP {\bf 1206}, 025 (2012)
  [arXiv:1203.0016 [hep-th]].



\bibitem{Tolley:2009fg} 
  A.~J.~Tolley and M.~Wyman,
  Phys.\ Rev.\ D {\bf 81}, 043502 (2010)
  [arXiv:0910.1853 [hep-th]].



\bibitem{Achucarro:2010jv} 
  A.~Achucarro, J.~-O.~Gong, S.~Hardeman, G.~A.~Palma and S.~P.~Patil,
  Phys.\ Rev.\ D {\bf 84}, 043502 (2011)
  [arXiv:1005.3848 [hep-th]].



\bibitem{Achucarro:2010da} 
  A.~Achucarro, J.~-O.~Gong, S.~Hardeman, G.~A.~Palma and S.~P.~Patil,
  JCAP {\bf 1101}, 030 (2011)
  [arXiv:1010.3693 [hep-ph]].



\bibitem{Achucarro:2012sm} 
  A.~Achucarro, J.~-O.~Gong, S.~Hardeman, G.~A.~Palma and S.~P.~Patil,
  JHEP {\bf 1205}, 066 (2012)
  [arXiv:1201.6342 [hep-th]].



\bibitem{Achucarro:2012yr} 
  A.~Achucarro, V.~Atal, S.~Cespedes, J.~-O.~Gong, G.~A.~Palma and S.~P.~Patil,
  Phys.\ Rev.\ D {\bf 86}, 121301 (2012)
  [arXiv:1205.0710 [hep-th]].



\bibitem{Achucarro:2012fd} 
  A.~Achúcarro, J.~-O.~Gong, G.~A.~Palma and S.~P.~Patil,
  Phys.\ Rev.\ D {\bf 87}, no. 12, 121301 (2013)
  [arXiv:1211.5619 [astro-ph.CO]].



\bibitem{Achucarro:2014msa} 
  A.~Achucarro, V.~Atal, B.~Hu, P.~Ortiz and J.~Torrado,
  arXiv:1404.7522 [astro-ph.CO].



\bibitem{Shiu:2011qw} 
  G.~Shiu and J.~Xu,
  Phys.\ Rev.\ D {\bf 84}, 103509 (2011)
  [arXiv:1108.0981 [hep-th]].



\bibitem{Cespedes:2012hu} 
  S.~Cespedes, V.~Atal and G.~A.~Palma,
  JCAP {\bf 1205}, 008 (2012)
  [arXiv:1201.4848 [hep-th]].



\bibitem{Gao:2012uq} 
  X.~Gao, D.~Langlois and S.~Mizuno,
  JCAP {\bf 1210}, 040 (2012)
  [arXiv:1205.5275 [hep-th]].



\bibitem{Konieczka:2014zja} 
  M.~Konieczka, R.~H.~Ribeiro and K.~Turzynski,
  arXiv:1401.6163 [astro-ph.CO].



\bibitem{Cremonini:2010ua} 
  S.~Cremonini, Z.~Lalak and K.~Turzynski,
  JCAP {\bf 1103}, 016 (2011)
  [arXiv:1010.3021 [hep-th]].



\bibitem{Peterson:2011yt} 
  C.~M.~Peterson and M.~Tegmark,
  Phys.\ Rev.\ D {\bf 87}, no. 10, 103507 (2013)
  [arXiv:1111.0927 [astro-ph.CO]].



\bibitem{Behbahani:2011it} 
  S.~R.~Behbahani, A.~Dymarsky, M.~Mirbabayi and L.~Senatore,
  JCAP {\bf 1212}, 036 (2012)
  [arXiv:1111.3373 [hep-th]].



\bibitem{Chen:2012ge} 
  X.~Chen and Y.~Wang,
  JCAP {\bf 1209}, 021 (2012)
  [arXiv:1205.0160 [hep-th]].



\bibitem{Pi:2012gf} 
  S.~Pi and M.~Sasaki,
  JCAP {\bf 1210}, 051 (2012)
  [arXiv:1205.0161 [hep-th]].



\bibitem{Gwyn:2012mw} 
  R.~Gwyn, G.~A.~Palma, M.~Sakellariadou and S.~Sypsas,
  JCAP {\bf 1304}, 004 (2013)
  [arXiv:1210.3020 [hep-th]].



\bibitem{Gao:2013zga} 
  X.~Gao,
  JCAP {\bf 1310}, 039 (2013)
  [arXiv:1307.2564 [hep-th]].



\bibitem{Emami:2013lma} 
  R.~Emami,
  JCAP {\bf 1404}, 031 (2014)
  [arXiv:1311.0184 [hep-th]].



\bibitem{Castillo:2013sfa} 
  E.~Castillo, B.~Koch and G.~Palma,
  arXiv:1312.3338 [hep-th].



\bibitem{Gao:2013ota} 
  X.~Gao, D.~Langlois and S.~Mizuno,
  JCAP10(2013)023
  [arXiv:1306.5680 [hep-th]].



\bibitem{Noumi:2013cfa} 
  T.~Noumi and M.~Yamaguchi,
  JCAP {\bf 1312}, 038 (2013)
  [arXiv:1307.7110 [hep-th]].



\bibitem{Easther:2013kla} 
  R.~Easther and R.~Flauger,
  JCAP {\bf 1402}, 037 (2014)
  [arXiv:1308.3736 [astro-ph.CO]].

\bibitem{Chen:2014joa} 
  X.~Chen and M.~H.~Namjoo,
  arXiv:1404.1536 [astro-ph.CO].

\bibitem{Bagger:1997dv} 
  J.~Bagger and I.~Giannakis,
  Phys.\ Rev.\ D {\bf 56}, 2317 (1997)
  [hep-th/9703202].



\bibitem{Watson:2004aq} 
  S.~Watson,
  Phys.\ Rev.\ D {\bf 70}, 066005 (2004)
  [hep-th/0404177].



\bibitem{Kofman:2004yc} 
  L.~Kofman, A.~D.~Linde, X.~Liu, A.~Maloney, L.~McAllister and E.~Silverstein,
  JHEP {\bf 0405}, 030 (2004)
  [hep-th/0403001].



\bibitem{Green:2009ds} 
  D.~Green, B.~Horn, L.~Senatore and E.~Silverstein,
  Phys.\ Rev.\ D {\bf 80}, 063533 (2009)
  [arXiv:0902.1006 [hep-th]].



\bibitem{Silverstein:2008sg} 
  E.~Silverstein and A.~Westphal,
  Phys.\ Rev.\ D {\bf 78}, 106003 (2008)
  [arXiv:0803.3085 [hep-th]].



\bibitem{Battefeld:2010sw} 
  D.~Battefeld and T.~Battefeld,
  JHEP {\bf 1007}, 063 (2010)
  [arXiv:1004.3551 [hep-th]].



\bibitem{Battefeld:2013bfl} 
  D.~Battefeld, T.~Battefeld and D.~Fiene,
  arXiv:1309.4082 [astro-ph.CO].



\bibitem{Langlois:2009jp} 
  D.~Langlois and L.~Sorbo,
  JCAP {\bf 0908}, 014 (2009)
  [arXiv:0906.1813 [astro-ph.CO]].



\bibitem{Battefeld:2011yj} 
  D.~Battefeld, T.~Battefeld, C.~Byrnes and D.~Langlois,
  JCAP {\bf 1108}, 025 (2011)
  [arXiv:1106.1891 [astro-ph.CO]].



\bibitem{D'Amico:2013iaa} 
  G.~D'Amico, R.~Gobbetti, M.~Kleban and M.~Schillo,
  JCAP {\bf 1311}, 013 (2013)
  [arXiv:1306.6872 [astro-ph.CO]].



\bibitem{Kofman:1997yn} 
  L.~Kofman, A.~D.~Linde and A.~A.~Starobinsky,
  Phys.\ Rev.\ D {\bf 56}, 3258 (1997)
  [hep-ph/9704452].



\bibitem{Battefeld:2006cn} 
  T.~Battefeld and N.~Shuhmaher,
  Phys.\ Rev.\ D {\bf 74}, 123501 (2006)
  [hep-th/0607061].



\bibitem{Battefeld:2012wa} 
  T.~Battefeld, A.~Eggemeier and J.~.Giblin, John T.,
  JCAP {\bf 1211}, 062 (2012)
  [arXiv:1209.3301 [astro-ph.CO]].



\bibitem{Chen:2008wn} 
  X.~Chen, R.~Easther and E.~A.~Lim,
  JCAP {\bf 0804}, 010 (2008)
  [arXiv:0801.3295 [astro-ph]].



\bibitem{Chen:2010xka} 
  X.~Chen,
  Adv.\ Astron.\  {\bf 2010}, 638979 (2010)
  [arXiv:1002.1416 [astro-ph.CO]].



\bibitem{Chen:2011zf} 
  X.~Chen,
  JCAP {\bf 1201}, 038 (2012)
  [arXiv:1104.1323 [hep-th]].



\bibitem{Chen:2011tu} 
  X.~Chen,
  Phys.\ Lett.\ B {\bf 706}, 111 (2011)
  [arXiv:1106.1635 [astro-ph.CO]].



\bibitem{Chen:2012ja} 
  X.~Chen and C.~Ringeval,
  JCAP {\bf 1208}, 014 (2012)
  [arXiv:1205.6085 [astro-ph.CO]].



\bibitem{Battefeld:2013xka} 
  T.~Battefeld, J.~C.~Niemeyer and D.~Vlaykov,
  JCAP {\bf 1305}, 006 (2013)
  [arXiv:1302.3877 [astro-ph.CO]].



\bibitem{Saito:2012pd} 
  R.~Saito, M.~Nakashima, Y.~-i.~Takamizu and J.~'i.~Yokoyama,
  JCAP {\bf 1211}, 036 (2012)
  [arXiv:1206.2164 [astro-ph.CO]].



\bibitem{Saito:2013aqa} 
  R.~Saito and Y.~-i.~Takamizu,
  JCAP {\bf 1306}, 031 (2013)
  [arXiv:1303.3839, arXiv:1303.3839 [astro-ph.CO]].



\bibitem{Kobayashi:2012kc} 
  T.~Kobayashi and J.~'i.~Yokoyama,
  JCAP {\bf 1302}, 005 (2013)
  [Erratum-ibid.\  {\bf 1309}, E02 (2013)]
  [arXiv:1210.4427 [astro-ph.CO]].



\bibitem{Noumi:2012vr} 
  T.~Noumi, M.~Yamaguchi and D.~Yokoyama,
  JHEP {\bf 1306}, 051 (2013)
  [arXiv:1211.1624 [hep-th]].



\bibitem{Mizuno:2014jja} 
  S.~Mizuno, R.~Saito and D.~Langlois,
  arXiv:1405.4257 [hep-th].



\bibitem{Joy:2007na} 
  M.~Joy, V.~Sahni and A.~A.~Starobinsky,
  Phys.\ Rev.\ D {\bf 77}, 023514 (2008)
  [arXiv:0711.1585 [astro-ph]].



\bibitem{Joy:2008qd} 
  M.~Joy, A.~Shafieloo, V.~Sahni and A.~A.~Starobinsky,
  JCAP {\bf 0906}, 028 (2009)
  [arXiv:0807.3334 [astro-ph]].



\bibitem{Starobinsky:1992ts} 
  A.~A.~Starobinsky,
  JETP Lett.\  {\bf 55}, 489 (1992)
  [Pisma Zh.\ Eksp.\ Teor.\ Fiz.\  {\bf 55}, 477 (1992)].



\bibitem{Adams:2001vc} 
  J.~A.~Adams, B.~Cresswell and R.~Easther,
  Phys.\ Rev.\ D {\bf 64}, 123514 (2001)
  [astro-ph/0102236].



\bibitem{Battefeld:2010rf} 
  D.~Battefeld, T.~Battefeld, H.~Firouzjahi and N.~Khosravi,
  JCAP {\bf 1007}, 009 (2010)
  [arXiv:1004.1417 [hep-th]].



\bibitem{Battefeld:2010vr} 
  D.~Battefeld, T.~Battefeld, J.~T.~Giblin, Jr. and E.~K.~Pease,
  JCAP {\bf 1102}, 024 (2011)
  [arXiv:1012.1372 [astro-ph.CO]].



\bibitem{Battefeld:2008py} 
  D.~Battefeld, T.~Battefeld and A.~-C.~Davis,
  JCAP {\bf 0810}, 032 (2008)
  [arXiv:0806.1953 [hep-th]].



\bibitem{Battefeld:2008qg} 
  D.~Battefeld and T.~Battefeld,
  JCAP {\bf 0903}, 027 (2009)
  [arXiv:0812.0367 [hep-th]].



\bibitem{GarciaBellido:1997te} 
  J.~Garcia-Bellido, J.~Garriga and X.~Montes,
  Phys.\ Rev.\ D {\bf 57}, 4669 (1998)
  [hep-ph/9711214].



\bibitem{Sugimura:2011tk} 
  K.~Sugimura, D.~Yamauchi and M.~Sasaki,
  JCAP {\bf 1201}, 027 (2012)
  [arXiv:1110.4773 [gr-qc]].



\bibitem{Firouzjahi:2014fda} 
  H.~Firouzjahi and M.~H.~Namjoo,
  arXiv:1404.2589 [astro-ph.CO].



\bibitem{Firouzjahi:2010ga} 
  H.~Firouzjahi and S.~Khoeini-Moghaddam,
  JCAP {\bf 1102}, 012 (2011)
  [arXiv:1011.4500 [hep-th]].



\bibitem{Namjoo:2012xs} 
  M.~H.~Namjoo, H.~Firouzjahi and M.~Sasaki,
  JCAP {\bf 1212}, 018 (2012)
  [arXiv:1207.3638 [hep-th]].



\bibitem{Israel:1966rt} 
  W.~Israel,
  Nuovo Cim.\ B {\bf 44S10}, 1 (1966)
  [Erratum-ibid.\ B {\bf 48}, 463 (1967)]
  [Nuovo Cim.\ B {\bf 44}, 1 (1966)].



\bibitem{Deruelle:1995kd} 
  N.~Deruelle and V.~F.~Mukhanov,
  Phys.\ Rev.\ D {\bf 52}, 5549 (1995)
  [gr-qc/9503050].



\bibitem{Martin:1997zd} 
  J.~Martin and D.~J.~Schwarz,
  Phys.\ Rev.\ D {\bf 57}, 3302 (1998)
  [gr-qc/9704049].



\bibitem{Linde:1993cn} 
  A.~D.~Linde,
  Phys.\ Rev.\ D {\bf 49}, 748 (1994)
  [astro-ph/9307002].



\bibitem{Copeland:1994vg} 
  E.~J.~Copeland, A.~R.~Liddle, D.~H.~Lyth, E.~D.~Stewart and D.~Wands,
  Phys.\ Rev.\ D {\bf 49}, 6410 (1994)
  [astro-ph/9401011].



\bibitem{Taruya:1997iv} 
  A.~Taruya and Y.~Nambu,
  Phys.\ Lett.\ B {\bf 428}, 37 (1998)
  [gr-qc/9709035].



\bibitem{Gordon:2000hv} 
  C.~Gordon, D.~Wands, B.~A.~Bassett and R.~Maartens,
  Phys.\ Rev.\ D {\bf 63}, 023506 (2001)
  [astro-ph/0009131].



\bibitem{Byrnes:2006fr} 
  C.~T.~Byrnes and D.~Wands,
  Phys.\ Rev.\ D {\bf 74}, 043529 (2006)
  [astro-ph/0605679].



\bibitem{Achucarro:2013cva} 
  A.~Achucarro, V.~Atal, P.~Ortiz and J.~Torrado,
  Phys.\ Rev.\ D {\bf 89}, 103006 (2014)
  [arXiv:1311.2552 [astro-ph.CO]].



\bibitem{Gong:2014spa} 
  J.~-O.~Gong, K.~Schalm and G.~Shiu,
  Phys.\ Rev.\ D {\bf 89}, 063540 (2014)
  [arXiv:1401.4402 [astro-ph.CO]].




\end{thebibliography}
\end{document}